\documentclass[preprint]{aastex}

\usepackage{emulateapj5}
\newcommand{\realfigure}[3]{ 
             \hbox{~} \centerline{\includegraphics[width=8.7cm]{#1}}
             \figcaption{#2 \label{#3}} \vspace{0.05in}\centerline{}}

\slugcomment{Version: 2004, September 9}

\shorttitle{Light to Mass Variations with Environment}
\shortauthors{Tully}

\received{2003 June 26; Revised 2003, Oct 6; 2004, Jan 23; 2004, March
  18; 2004, July 3; 2004, September 9}
\begin{document}


\title{Light to Mass Variations with Environment}


\author{R. Brent Tully}
\affil{Institute for Astronomy, University of Hawaii, Honolulu, HI 96822}

\begin{abstract}
Large and well defined variations exist between the distribution of mass
and the light of stars on extragalactic scales.  Mass concentrations in the
range $10^{12} - 10^{13} M_{\odot}$ manifest the most light per unit mass.
Group halos in this range are typically the hosts of spiral and irregular 
galaxies with ongoing star formation.  On average
$M/L_B \sim 90 M_{\odot}/L_{\odot}$ in these groups .  More massive halos 
have less light per 
unit mass.  Within a given mass range, halos that are dynamically old as
measured by crossing times and galaxy morphologies have distinctly less
light per unit mass.  At the other end of the mass spectrum, below 
$10^{12} M_{\odot}$, there is a cutoff in the manifestation of light.
Group halos in the range $10^{11} - 10^{12} M_{\odot}$ can host dwarf galaxies
but with such low luminosities that $M/L_B$ values can range from several hundred
to several thousand.  It is suspected that there must be completely dark halos
at lower masses.  Given the form of the halo mass function, it is the low
relative luminosities of the high mass halos that has the greatest cosmological
implications.  Of order half the clustered mass may reside in halos 
with greater than $10^{14} M_{\odot}$.  By contrast, only $5-10\%$ of clustered
mass would lie in entities with less than $10^{12} M_{\odot}$.
\end{abstract}



\section{Introduction}

It would be remarkable if the naive assumption that light linearly traces 
mass is correct.  Indeed, on scales of tens of kiloparsecs around galaxies
it is known from rotation curve information that dark matter is more 
dispersed than starlight.  It has long been entertained that dark matter
might be more dispersed than light on megaparsec scales 
\citep{kai84}.  If so, we would say there is a `bias' between what
we see and the physically more fundamental parameter, the mass.  
Semi-analytic models \citep{bla99,som01,ost03} anticipate variations in the
conversion of baryons into stars and the manifestation as starlight
at both very low and very high densities.

Theoretical considerations prepare us for the possibility of a complex
relationship between the distribution of mass and of the lighthouses 
sitting atop that mass.  In this article we present 
observational evidence for such complexity.  Three environmental regimes
will be considered.  At home in the Local Group we find ourselves in an
intermediate regime, common to the majority of spiral galaxies.
Most elliptical galaxies find themselves in a high density regime.
Then there is a low density regime where galaxies are few or absent.
What is the relationship between light and mass in each of these three
kinds of environments?

\section{Groups of Galaxies}

Visible galaxies are embedded in halos that extend beyond the radii of
observable gas and stars.  We need to go to the scale of groups and clusters
to obtain a measure of the total mass of galaxies.  The current investigation relies
on the analysis of groups by \citet{tul87}; hereafter T87.  The individual 
galaxies associated
with the groups in T87 are identified in Table II of \citet{tul88}.  This particular
group compendium has the attractive features that (a) it was constructed with
quantitative, well defined rules, (b) the groups have been demonstrated to contain 
few interlopers, and (c) a subset has a high level of volume completeness.

The group catalog was constructed through a tree or `dendogram' procedure.
The construction begins by looking for the linkage between galaxies that gives
the largest value of the product $L_B/R_{ij}^2$ where
two candidates $i$ and $j$ are considered from the ensemble of $N$ objects,
$L_B$ is the blue luminosity of the brighter candidate, and $R_{ij}$ is the 
linear separation between the two.  For details on the calculation of $R_{ij}$
from angular and radial velocity information see T87 but essentially,
for linkages relevant to group scales, the $R_{ij}$ are based on angular 
separations only.  The dominant pair that is selected is now inserted back
into the catalog as a single unit with the sum of the component luminosities 
and with position and velocity constructed from the luminosity weighted
contributions of the components.  There are now $N-1$ objects to consider and
the process is repeated, over and over, until $N=1$.

In a dendogram constructed in this manner, ultimately all galaxies under
consideration are linked.  Obviously there is a transition at some point
from collections of galaxies that are bound to those that are unbound.
The issue of how to distinguish the bound and unbound domains was the
subject of extensive discussion in T87.  A crossing time argument was used.
At any point in the dendogram, the galaxies that are linked together can be
characterized by a separation dimension, the inertial radius $R_I$, and a 
velocity dispersion,
$V_p$.  The crossing time $t_{\rm x} \sim R_I / V_p$ (see T87 for details 
including
corrections for projection) can be compared with the free expansion timescale,
the inverse of the Hubble Constant, H$_0$.  Units with 
$t_{\rm x} {\rm H}_0 << 1$ have a high probability of being bound.  Units with
$t_{\rm x} {\rm H}_0 > 1$ can be inferred to be unbound.  With a
judicious choice of level, the dendogram could be split 
between mostly bound and mostly unbound.  It would follow from this choice,
allowing for measurement and projection errors, that essentially all the
merged units that make the dendogram cut on the high density side would have 
$t_{\rm x} {\rm H}_0 < 1$.

An interesting aspect of this analysis is that it provides an inventory not 
only of all the galaxies in a volume that are in groups but also {\it all 
that are not} in groups.  In the sample considered by T87 there was 
the usual incompletion with distance.  However a high Galactic
latitude volume with the distance limit of $25 h_{75}^{-1}$ Mpc
($h_{75} = {\rm H}_0/75$) has completion 
in the sense that all galaxies
brighter than $M_B^{\star}$ are included.  Within this volume, T87 identified 
179 groups of 2 or more galaxies and 49 groups of 5 or more.  For statistical
robustness, we will restrict most of the following discussion to the groups 
of 5 or 
more galaxies (one of the small groups in T87 has been boosted to above 5 
members with our current inventory,
giving us 50 groups to consider).  

We should note a subtle point that arises because the dendogram linkages
involve luminosities.  Linkages that involve bright galaxies are made more 
easily, as might well be reasonable if brighter galaxies are more massive.
However could we miss linkages in cases where mass is underrepresented by 
light?  This possibility would potentially undermine the basis of this
investigation.  It is to be noted that T87 provides a safety net
to guard against this concern. In addition to the dendogram cut that resulted
in the definition of `groups', a second cut was made an order of magnitude 
lower in luminosity density that resulted in what were called `associations'.  
We will
first continue with a discussion of the groups but later we will revisit
the entities called associations.

\section{Light to Mass Variations within Groups}

The analysis in T87 missed an amazing correlation!  It is seen in Figure 1,
which draws on data extracted from T87 and presented here in Table 1.
The data in the left panel represent all the 50 groups in the volume with
completion above $M_B^{\star}$ .  In the right panel, symbols distinguish
between groups with majority early morphological types (E-S0-Sa: filled 
squares) and groups with majority late types (Sab-Irr: open circles).
There are fewer points in this panel because, in order to make a statistically
meaningful distinction in morphological class, we require there be at least 
6 galaxies brighter than $M_B=-17$.  These groups are distinguished by an entry
in the third column of Table 1 where the percentage type Sa or 
earlier among galaxies brighter than $M_B=-17$ is given.
The numbers of early type groups
are small in this restricted local volume, so the slightly more distant
early type groups Virgo~W, Antlia, and NGC 5846 are
also considered (last 3 entries in Table 1).

Groups with crossing times much less than a Hubble time have relatively 
high mass to light ($M/L_B$) ratios, values of several hundred in solar units
(masses from T87 are unweighted virial masses following the definition
given in Section 6 of this paper).
Groups with crossing times that approach a Hubble time have $M/L_B$ ratios
that are much more modest, a few tens in solar units.  It is also seen
that there
is a strong correlation between crossing time and the morphological 
classification.  If the sample is split at a crossing time as a fraction of
the Hubble timescale of 
$0.2 {\rm H}_0$ then almost all the groups with shorter crossing times
are dominated by early types (8 of 9) and almost all the groups with longer
crossing times are dominated by late types (11 of 14).

\realfigure{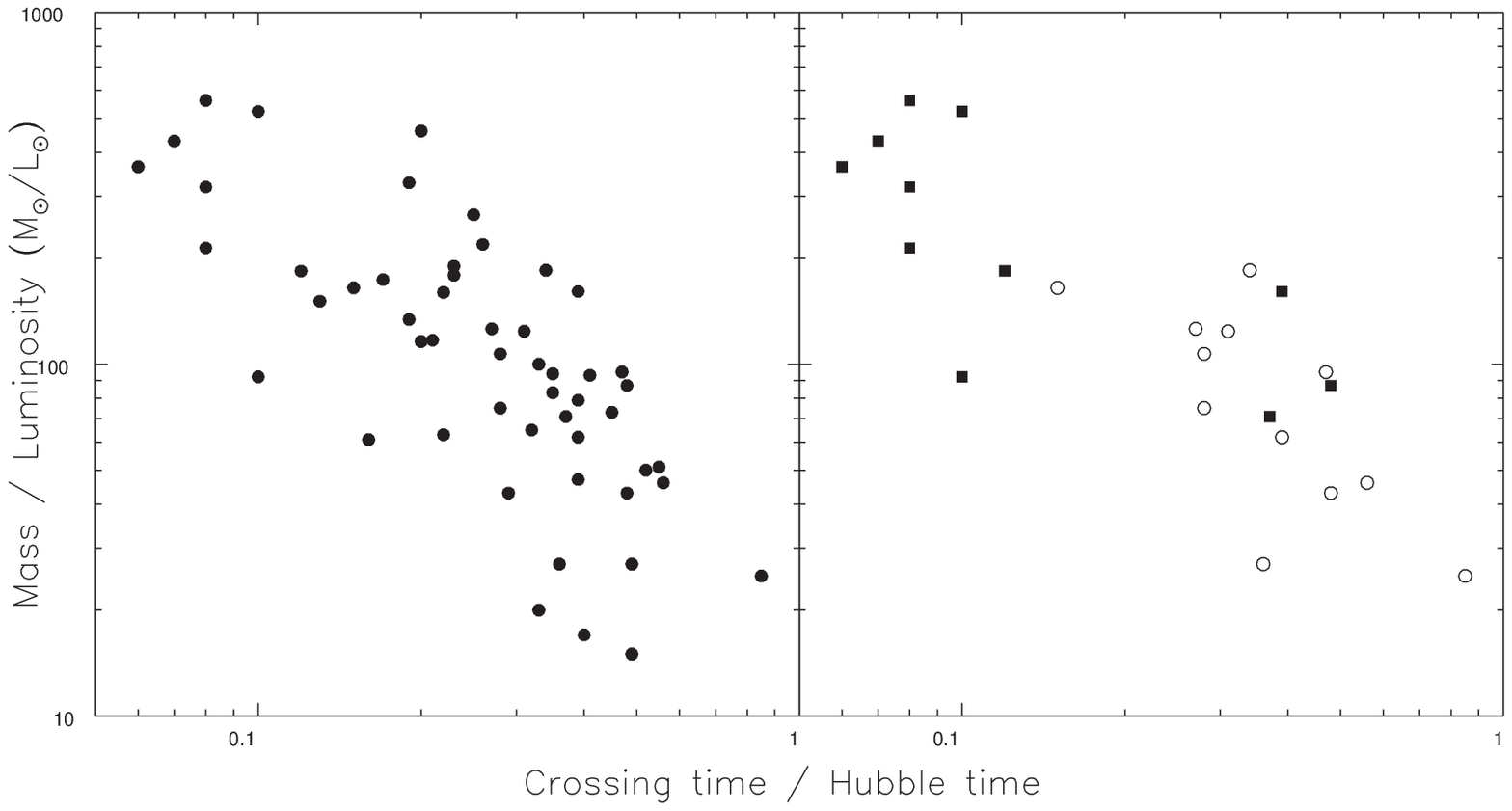}{Correlation between mass/light ratio and group crossing time.
{\it Left panel:} all groups with 5 or more members in the volume limited 
T87 sample.  {\it Right panel:} Subset of groups with at least 6 members
with $M_B<-17$ distinguished according to whether a majority of
galaxies in the group are E-S0-Sa
(filled squares) or Sab-Sc-Irr (open circles).}{fig1}


Could the correlation seen in Fig.~1 be induced by parameter coupling?
The vertical axis has dependencies on $V_p^2 R$ while the horizontal axis 
has dependencies $R/V_p$, where $R$ is a characteristic of the group dimension
and $V_p$ is the velocity dispersion.  Changes in $V_p$ cause shifts in the 
sense of the observed correlation.  Changes in $R$ cause shifts orthogonal to
the correlation.  The concern would have to be with $V_p$ values but the
typical measurement uncertainties in these values are 20\% for the groups in
question, much smaller than the factor ten range in crossing time values.
Most clearly, though, the correlation cannot be an artifact of
parameter errors and coupling between the axes
because there is a strong correlation between the parameter of either 
axis in Fig.~1 and the morphology of the galaxies.  The qualitative
morphological information is decoupled from the measured parameters.  In the
right panel of Fig.~1, the early type groups have a median crossing time
about a factor 3 less and a median $M/L_B$ about a factor of 3 higher than the 
late type groups.

Could the correlation be an artifact caused by systematic departures
from virial equilibrium?  Presumably the groups with the shortest
crossing times have most closely approached the virial equipartition
of energy.  There could be a systematic mis-estimation of mass with
the virial approximation that depends on the dynamical age of a
system.  
Pacheco (private communication) has studied the collapse of a group
with the expression:
\begin{equation}
{1 \over 2} {{\rm d}^2 I \over {\rm d} t^2} = {2 (T+U) + W }
\end{equation}
where $I$ is the moment of inertia of the system, $T$ is the kinetic
energy in the randomized motions of the constituents, $U$ is the
kinetic energy in the ordered motions of infall, and $W$ is
the potential energy.  Pacheco made analytic assumptions about what
happens when, during the collapse,
violent relaxation \citep{lyn67,fun92} causes the transform of energy
from bulk motion to a state approximated by the virial condition.
Dynamical friction induces the transfer of energy from ordered to
random motions.  The approach to equilibrium, with 
${\rm d}^2 I /{\rm d} t^2 \sim 0$, occurs on the order of one
dynamical timescale ($t_{dyn} = (G\rho)^{-1/2} \sim t_x$ where $\rho$ is
the average density of the group within the virial radius).
Measurements of groups caught before relaxation would tend to result
in underestimates of mass by typically 20\% and a factor 2 in the
extreme (theoretically assuming full 3-D information is
available).  

The situation is complex.  It was suggested by \citet{bar85} that
virial mass estimates might be biased low because of segregation
as the visible components sink toward the center relative to dark
matter.   This possible systematic would increase as the group ages,
consequently it would run counter to the correlation seen in Fig.~1.
There is room for further analysis of the properties of groups
extracted from cosmological simulations.  High resolution is required
to avoid the concern of artificial 2-body relaxation from massive
numerical particles. 

There is firmer ground on the observational side.  There is general
agreement that groups and clusters with high densities and short
dynamical times have $M/L_B$ values of several hundred.  These
circumstances are confirmed by X-ray and gravitational lensing
studies.  We suggest that the trend to substantially lower $M/L_B$
values for systems with longer dynamical times seen in Fig.~1 is
consistent with what is now known from the few detailed studies of low
density groups.

The situation is clearest in the Local Group and has received
confirmation from recent studies of the nearest neighboring groups.
These groups are represented in the left panel of Fig.~1 but are not
sufficiently populated to enter the right panel.  They lie in the
lower right corner, with large crossing times and low $M/L_B$.

The Local Group encompasses two dynamical regimes.  On a 100~kpc scale,
each of the two major galaxies host a school of minor systems that
approximate virial conditions.  On a 1~Mpc scale, the two major
galaxies and ten or so other small galaxies are falling together, most
on first approach.  This latter assertion is based on the observation
that almost all the Local Group systems that are not tightly
associated with either of the two major galaxies have negative
velocities in the Local Group standard of rest (NGC 6822 is an
exception).
 
The masses of the Milky Way and Andromeda sub-groups can be estimated
through the virial theorem since these sub-systems have densities and
crossing times that suggest they are dynamically evolved \citep{eva00}.
The stellar stream on a scale of 125 kpc around M31 provides a
compatible mass estimate \citep{iba04}.  Each sub-group is determined
to have $0.7-1 \times 10^{12} M_{\odot}$.
The mass of the larger infall region can be estimated separately.  One
approach follows the \citet{kah59} timing argument.  A variation on this
approach is provided by the identification of the current turnaround
radius or zero-velocity surface for the Local Group \citep{san86, kar02c}.
There is the important result that the sum of the masses of the
Andromeda and Milky Way sub-systems add to approximately the mass
determined to lie within the zero-velocity surface of 
$1.2 \times 10^{12}  M_{\odot}$.  A similar mass is
found from orbit reconstructions \citep{pee95}.  Our naive virial
estimation (Table 1: group 14-12) gives a Local Group mass of 
$1 \times 10^{12} M_{\odot}$ and $M/L_B=17 M_{\odot}/L_{\odot}$, in
factor 2 agreement with other results.

Recent observations of the resolved stars in nearby galaxies with
Hubble Space Telescope have dramatically improved our knowledge of
distances to many nearby galaxies and hence of the structure of the
nearest groups.  We now know that the best studied neighboring groups,
those about M81 (Table 1: group 14-10) and Centaurus~A (Table 1: group
14-15), have dumbbell structures like the Local Group.  In both cases 
the two largest galaxies are surrounded by swarms
of small galaxies which transit their hosts in times much less than
the age of the universe.  Almost certainly in the case of the M81
Group and plausibly in the case of the Centaurus Group, the
substructures are falling together.  As with the Local Group, the
masses inferred for the entire bound entities 
(2 and 3 times $10^{12} M_{\odot}$ respectively) are close to the sum
of the masses of the dynamically evolved subcomponents
\citep{kar02a,kar02b}.  The halos of 
the giant galaxies must not extend much beyond the domain of the
immediate companions and the group $M/L_B$ values are a few tens. 
These groups that are overall spiral rich
and low density evidently have $M/L_B$ values well below 100.  The virial
approximation for these groups, while crude, gives mass estimates that
are not strongly deviant.

While theoretical expectations might cause one to anticipate
systematic underestimations of mass assuming the virial approximation,
studies of the nearest structures reveal that a crude application of
the virial theorem may frequently result in {\it overestimations} of
mass.  Frequently, unbound objects will be erroneously included as
group members. The availability of good distances to individual
galaxies shows that the very nearby Sculptor Group (Table 1: 14-13)
and CVn I Group (Table 1: 14-07) are composites of regions that
are unlikely to be bound in the ensemble.  The inclusion of expansion
velocities and the inflation of group scale cause overestimates of
mass.  These effects are pernicious for the extended, low dispersion
systems that inhabit the lower right corner of Fig.~1.

The limited information available on morphological variations
from studies of nearby groups is giving a hint of higher mass in early
type systems.  The kinematics of the sub-group around Centaurus~A
suggest a mass of $4 \times 10^{12} M_{\odot}$ and 
$M/L_B \sim 100 M_{\odot}/L_{\odot}$ (albeit including the large
spiral NGC 4945) while the M83 sub-group is found
to have a mass of  $1 \times 10^{12} M_{\odot}$ and 
$M/L_B \sim 50 M_{\odot}/L_{\odot}$.  The elliptical Cen~A may have
a higher $M/L_B$ by a factor 2--3 relative to the large spirals in our
vicinity.  The situation in the nearest
groups is reviewed by \citet{kar04}.

Figure 2 shows the correlation between mass and light for the TF87
sample of nearby groups.
In the left panel, all the 50 groups in the volume-limited sample with at least
5 members are represented.
The groups with $t_{\rm x} {\rm H}_0 < 0.2$ are plotted with filled squares
and the groups with larger crossing times are plotted with open circles.
The 3 more distant early type groups, all with $t_{\rm x} {\rm H}_0 < 0.2$,
are plotted as inverted triangles.  The solid line at $45^{\circ}$ indicates 
the mean $M/L_B$ value determined by T87
for this sample.  Again we see that the groups with short crossing times have
less blue luminosity per unit of mass.  In addition, it can be said that groups
with more mass manifest less blue light per unit mass.

We consider the same data in the right panel but now the differentiation is made
on the basis of the morphology of the groups.  It is seen again that the
qualitative morphological description and the quantitative crossing time
measure provide an equivalent basis for distinguishing between high and low
$M/L_B$ systems.

\realfigure{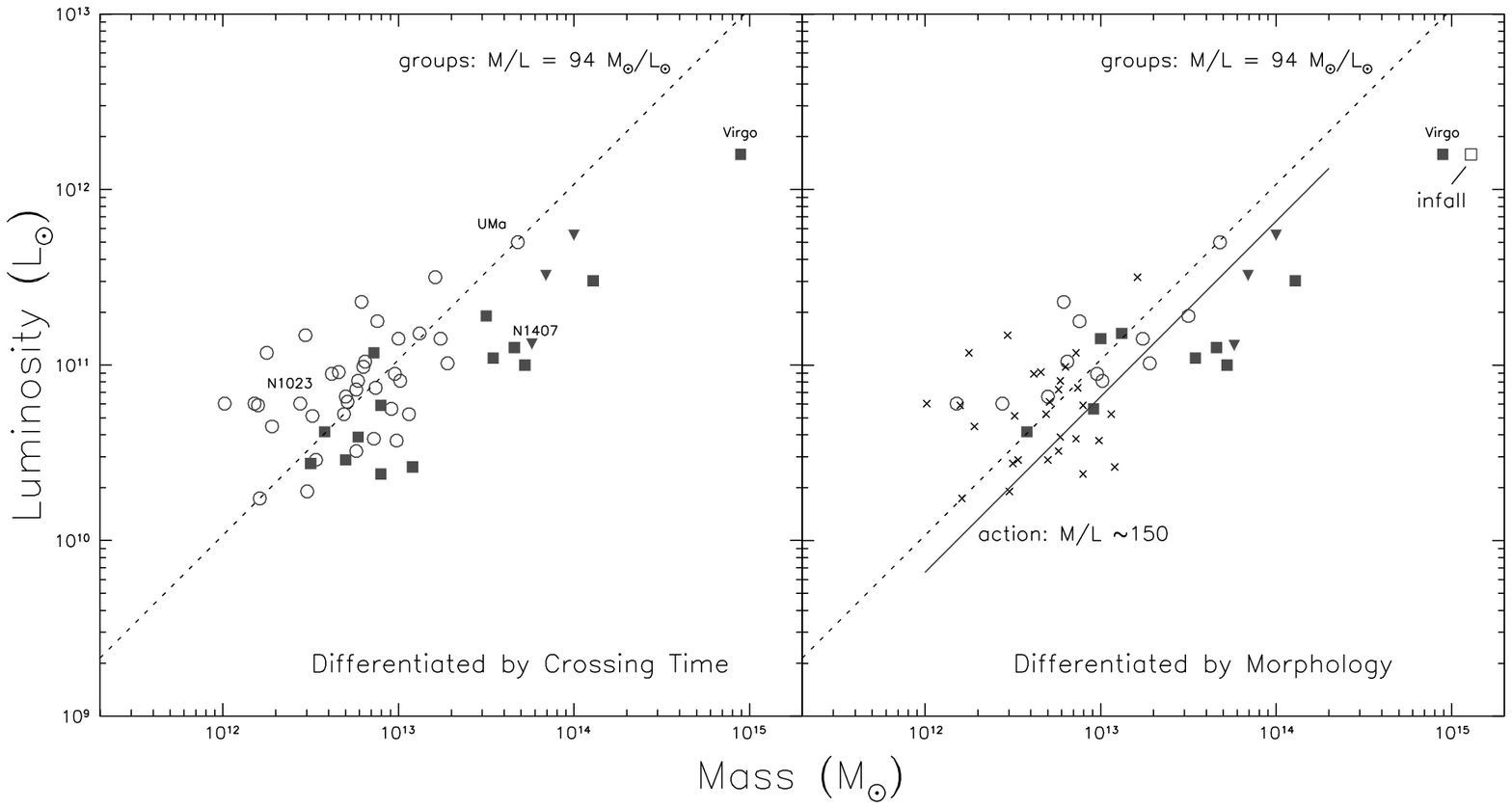}{Correlation of mass and blue luminosity.  {\it Left panel:} 
Groups with $t_{\rm x} {\rm H}_0 < 0.2$ are indicated by filled squares and
inverted triangles and
groups with longer crossing times are indicated by open circles.  The mean
$M/L_B$ value of $94 M_{\odot}/L_{\odot}$ for T87 groups is shown by the 
$45^{\circ}$ dotted line.  The 4 representative groups that are identified
(Virgo, Ursa Major, NGC 1023, NGC 1407) are discussed in the text.  The
3 inverted triangles represent groups at larger distance than the complete sample.
{\it Right panel:} Same data points but now majority E-S0-Sa groups are 
indicated by filled squares and inverted triangles and majority Sab-Sc-Irr 
groups are indicated by 
open circles.  Groups too sparse to be differentiated by morphology are 
indicated by crosses.  This figure shows the $M/L_B=94 M_{\odot}/L_{\odot}$
dotted line and, in addition, shows the mean result for the field that comes
from Numerical Action models, the $M/L_B=150 M_{\odot}/L_{\odot}$ line.
Also, the mass of the Virgo Cluster given by infall constraints in the 
Numerical Action models is illustrated by the position of the open square.}{fig2}


The Virgo Cluster (group 11-1 in TF87) and three other specific groups are 
identified in the left 
panel of Fig.~2.  It is instructive to consider the pairwise properties
of these groups.  It can be seen that the NGC 1407 (51-8) Group has roughly the
same mass as the Ursa Major Cluster (group 12-1) but much lower
luminosity.  The NGC~1407 Group has
the same luminosity as the NGC 1023 (17-1) Group but much more mass.  These 
similarities and differences are not in observational doubt.  There is a 
summary of the properties of these groups in Table 1 of \citet{tre02}; see
also Fig.~1 in that reference for histograms of the group velocities.
The unusual nature of the NGC 1407 Group has been noted by \citet{gou93}
and \citet{qui94}.  The group contains only two $L^{\star}$ galaxies but has 
a velocity dispersion of 385 km~s$^{-1}$.  The Ursa Major Cluster, by contrast,
has almost enough luminous galaxies to qualify as an Abell richness class 0
cluster!  The velocity dispersion, though, is only 148 km~s$^{-1}$.  The virial
masses of the two groups are almost the same.  In the NGC~1407 Group 
motions are large in a small volume while in the Ursa Major Cluster motions are
low in a large volume.  Yet the luminosities differ by a factor 5 and
$M/L_B$ values differ accordingly.
The NGC~1023 Group is a scaled down version of the Ursa Major Cluster.
There are three $L^{\star}$ galaxies and the group velocity dispersion is only
57 km~s$^{-1}$.  The NGC~1023 and NGC~1407 groups have similar galaxy
content in terms of numbers and luminosity (the type content is late and 
early, respectively).  However the motions in the NGC~1407 Group are a
factor of 7 higher.  This difference can not be questioned observationally, 
and the implied mass of the NGC~1407 Group is higher by a factor 30.

In summary of this section, evidence is provided of an order of
magnitude variation of $M/L_B$ that depends on mass and/or stage of
dynamical evolution.  The derivation of mass from the virial theorem
might have biases but there are effects that both raise and lower
estimates. The cases that are particularly well studied provide
confirmation of the trends shown in Figs.~1 and 2.

\section{Numerical Action Models of Large Scale Flows}

There is a strong inference from Numerical Action orbit reconstructions that
the Virgo Cluster is very underluminous for its mass compared with the general
field.  From a first analysis of the Local Supercluster \citep{sha95} with
300 distance estimates as constraints and a later analysis \citep{tul98} with
900 distance estimates, it was determined that the mean density of the Universe
is $\Omega_m \sim 0.2$, consistent with an overall mean 
$M/L_B \sim 200 M_{\odot}/L{_\odot}$ assignment.
However, as emphasized in the latter paper, it is not possible with a 
{\it single} assignment of $M/L_B$ for all galaxies to obtain a satisfactory 
model that simultaneously gives a good description of flows in the general
vicinity of the Local Supercluster and a good description of the infall
region around the Virgo Cluster.

Evidence for this claim is provided in Figures 3 and 4.  The first of these
figures show velocity-distance data for galaxies in a group along a specific
line-of-sight with respect to the Virgo Cluster.  The two panels illustrate 
attempts to fit this data with two distinct Numerical Action models.  The
solid curves represent the run of velocity with distance expected by the
separate models.  In the left panel, a value $M/L_B=200 M_{\odot}/L_{\odot}$
is assigned to all groups and galaxies.  The location of the galaxy data points
in velocity and distance cannot be understood in the context of the model in
this panel.  In the right panel, $M/L_B=1000 M_{\odot}/L_{\odot}$ is assigned
to the Virgo Cluster, and incidentally to all E/S0 knots.  In compensation
for this increase in the mass of some of the objects, 
$M/L_B=150 M_{\odot}/L_{\odot}$ is given all the rest.  Now the infall 
motions toward Virgo
are greatly enhanced and the model gives a physical basis for the
location of this particular group in velocity-distance space.  We see that
the group must be slightly nearer than the cluster and falling away from us
into the cluster.

The swing in amplitude of the wave in velocity-distance seen in Fig.~3 
(the `triple-value' characteristic \citep{ton81}) depends on the
mass assigned to the cluster.  It can be seen that the curve in the right
panel of Fig.~3 only minimally reaches the location of the data points, hence
represents a {\it minimum} required mass.  A larger mass would cause a
larger swing and would not be in 
conflict with the data but is not required.  Fig.~4 shows
the constraints provided by many lines-of-sight through the Virgo infall 
region.  The individual points record velocity and angular distance from 
Virgo, as open circles for galaxies in the cluster proper (defined by 
the `caustic' radius established by galaxies that have fallen into the cluster 
and back out to a second turnaround), and as filled circles for galaxies
identified to be within the first infall region.  To keep the plot clean,
galaxies at foreground and background `triple-value' locations have been 
rejected based on distance information.
The swings in amplitude of the triple-value waves along various lines-of-sight
are indicated by the brackets in the two panels.  It is clear that the swing
generated by the Numerical Action model of the left panel, with 
$M/L_B=200 M_{\odot}/L_{\odot}$ for all points, fails by a wide margin to 
explain the infall motions into the Virgo Cluster.  The model used in the
right panel, with $M/L_B=1000 M_{\odot}/L_{\odot}$ given to the Virgo Cluster
and $M/L_B=150 M_{\odot}/L_{\odot}$ assigned to the field, provides a 
satisfactory description to the observed data.

\realfigure{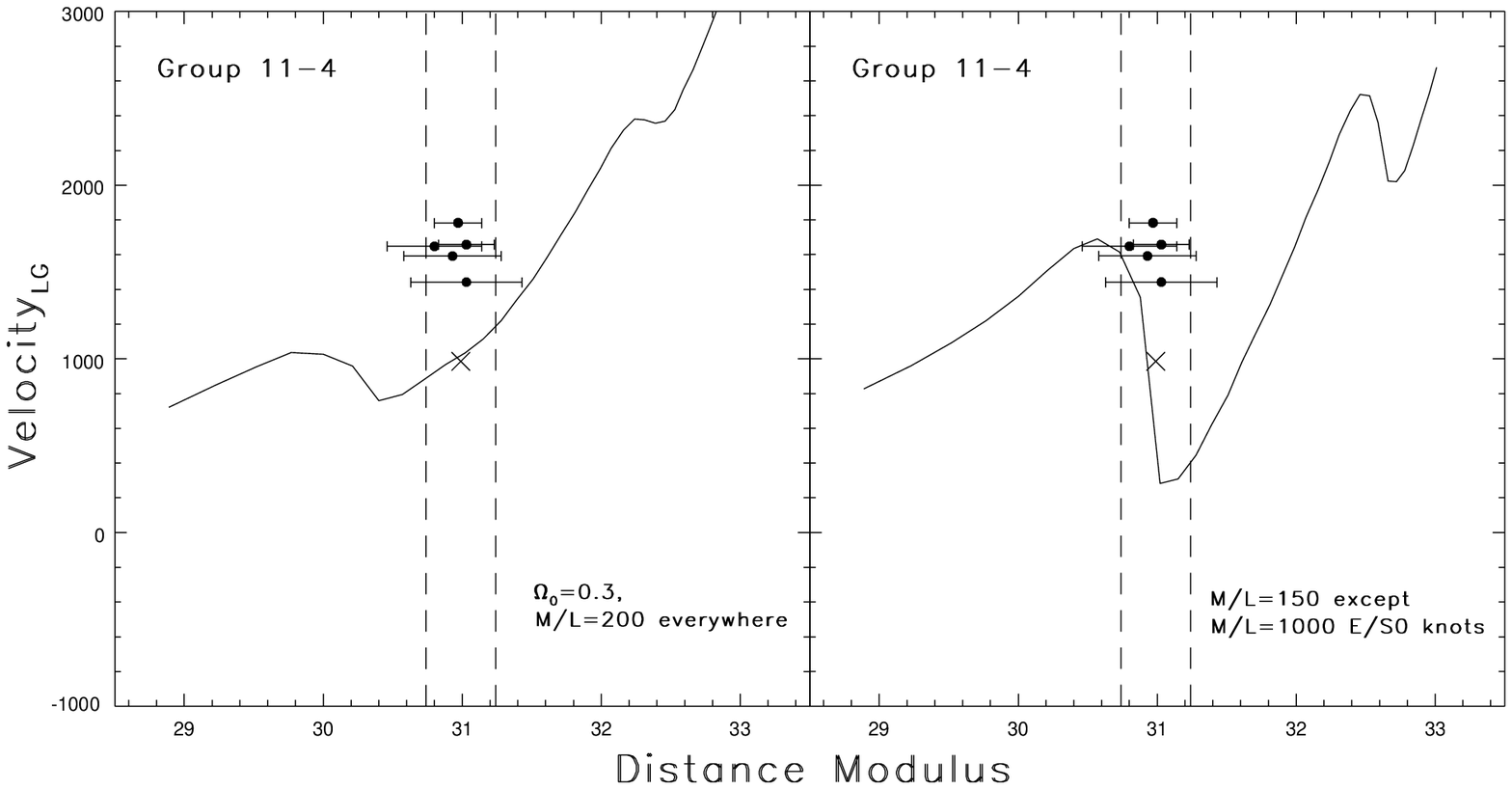}{Example of model velocities along a line-of-sight close to the
direction of the Virgo Cluster.  In this case, the line-of-sight is through
Group 11-4 in the catalog of \citet{tul88}.  The points with errors 
correspond to galaxies in this group with distance determinations.  The 
vertical dashed lines bracket the distance of the Virgo Cluster, centered
in distance and velocity at the large cross.  {\it Left panel:} $M/L_B=200$
for all entries.  {\it Right panel:} $M/L_B=1000$ for Virgo and other E/S0 
knots, otherwise $M/L_B=150$.  The curves are locii of velocities
allowed by the models as a function of distance in the specified
line-of-sight. The second wave in the velocity curve beyond 
Virgo occurs because the line-of-sight passes near another E/S0 knot,
the Virgo~W Cluster, Group 11-24.}{fig3}


\realfigure{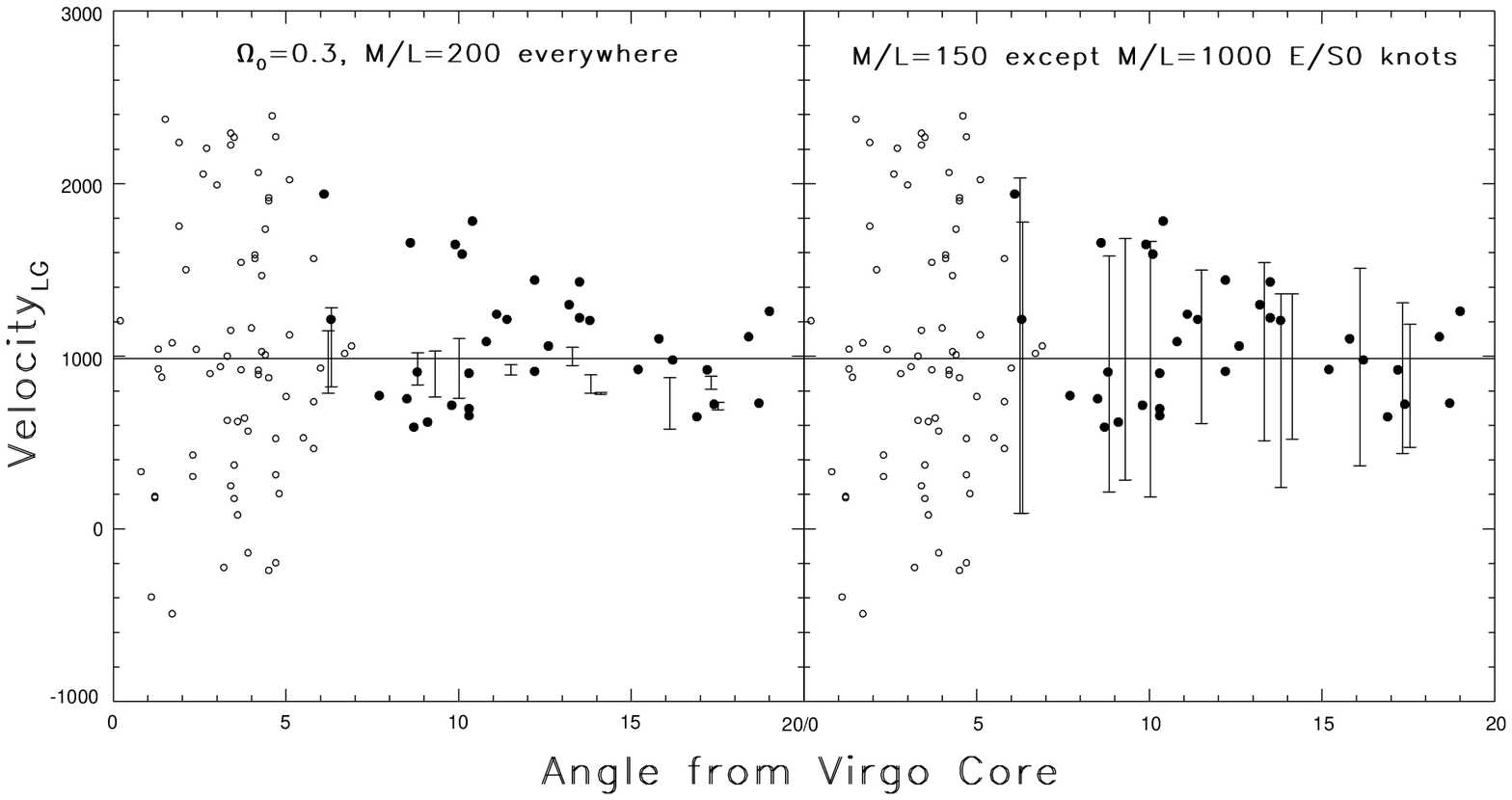}{Virgo infall constraints from two Numerical Action models.
Data points indicate the velocities and separations from the center of the 
Virgo Cluster of individual galaxies.  Galaxies within the $6^{\circ}$
caustic of the cluster are indicated by open circles.  Galaxies outside the
caustic but identified with the infall region are indicated by filled circles.
The vertical brackets are located at angles from the center of the cluster
that intersect infalling groups, so at lines-of-sight that have received
attention in the Numerical Action models.  The amplitudes of the brackets
illustrate the range of infall velocities anticipated by the two models
under consideration. {\it Left panel:} $M/L_B=200$ assigned to all units.
{\it Right panel:} $M/L_B=1000$ assigned to Virgo Cluster and other E/S0 knots
and $M/L_B=150$ otherwise.}{fig4}


{\it Almost a factor 7 higher $M/L_B$ value is required for the Virgo Cluster
than for the general field.}  The evidence for a lot of mass in the cluster
which comes from the extended nature of the infall region (extending to 
$\sim25^{\circ}$
from the center of the cluster or 8 Mpc) and high infall velocities (galaxies
hit the caustic surface of the cluster at $6^{\circ}$ with a velocity of 
1500 km~s$^{-1}$) requires a remarkable $1.3 \times 10^{15} M_{\odot}$ within
the 2 Mpc caustic surface of second turnaround.  This mass is almost
twice the virial mass estimate 
but it pertains to a radius three times larger than the virial radius.
The infall mass estimate pertains to the cluster on it's largest scale.
It can be noted that a similar conclusion was reached much earlier 
\citep{tul84} with a non-linear but spherically symmetric analytic model
of Virgo Cluster infall.  The very same conclusions were reached that there
had to be a lot of mass for the amount of light in the cluster but much less 
mass with respect to
light in the supercluster environs surrounding the cluster.

The preferred model in the \citet{tul98} analysis used two discrete $M/L_B$
assignments: $M/L_B=1000 M_{\odot}/L_{\odot}$ was assigned to 12 E-S0 dominated
groups including the Virgo Cluster and $M/L_B=150 M_{\odot}/L_{\odot}$ was
assigned otherwise.  
(The action models do not provide constraints on the masses of the E-S0 groups
smaller than Virgo because the triple value regions are too restricted to
map infall.  The large $M/L_B$ assignments for these groups were based on the
virial information discussed earlier.)
It was concluded that $\Omega_m \sim 0.2$ associated
with galaxies in the mean but that the Virgo Cluster in particular must have
a much higher $M/L_B$ value than the mean.  The Numerical Action mean result 
for the field ($M/L_B=150 M_{\odot}/L_{\odot}$) and 
the large mass for Virgo ($M \sim 1.3 \times 10^{13} M_{\odot}$) are recorded 
in the right panel of Fig.~2.

As an aside, it should be noted that what is meant by the terms `group' and
`cluster' are not well defined, other than that there is the inference
that the structures are bound.
In the case of 
the Local Group, it was discussed
that there is a zero-velocity surface at $\sim 1$~Mpc radius
\citep{san86}.  This surface defines a bound domain and provides a logical
limit to the group.  However, the Local Group is certainly not virialized
over this domain.  At most, there may be limited quasi-virialized regimes
within $\sim 150$~kpc about each of the two dominant galaxies.  By contrast,
when we talk about the Virgo Cluster we certainly do not consider it to 
extend to the current zero-velocity surface which lies on the near side at
about half the distance from us to the cluster core.  Rather, a good
working definition 
of the volume of the cluster is the domain within the first caustic, the outer
limit of objects that have completed a single passage.  The cluster is not 
virialized to this radius but the approach toward virialization will be
greater within a cluster defined this way than in the Local Group analog.

The discussion to this point, involving the properties of groups and the
inferences drawn from modeling of flow patterns in the Local Supercluster,
has dealt with comparisons between intermediate and high density regimes.
Attention will now turn to lower density environments.

\section{The Low Density Regime}

The algorithm for the selection of groups used by T87 contained luminosity.
Could we loose real groups because their constituents are underluminous?
We have just been arguing that groups with short crossing times and dominated
by early type galaxies are environments that are underluminous compared with
the environments of groups with crossing times approaching a Hubble time and
dominated by spirals and irregulars.  Fortunately, no groups of the former
type would have been missed in the construction of the group catalog because
the E/S0 groups with short crossing times have considerably higher 
spatial densities 
than the spiral groups.  Consequently they comfortably cross the luminosity
density threshold set for the group catalog even though they turn out to be
underluminous.  In the years that have passed since the catalog was developed
we have never become aware of a high spatial density but low luminosity
group in the volume of our sample that escaped identification.

However we cannot be so sanguine in the limit of low spatial densities.  Indeed, 
among the entities called `associations' that came out of the T87 study
there were some that look suspiciously like the entities identified as groups.
The dimensions of these units are only several hundred kiloparsecs, like the
groups.  They can have {\it very low} velocity dispersions, comparable with
the lowest values in the established groups.  {\it Their members are all 
dwarfs.}  They were not picked up in the groups analysis because they have
very low luminosities.  If they are bound groups then the associated $M/L_B$ is
much larger than is familiar to us from our experience with the intermediate
regime spiral groups.

Our interest in these entities was rekindled by work from a very different
direction.  We were interested in possible environmental dependencies at the
faint end of the luminosity function of galaxies.  It seems that there are
unexpectedly few dwarf galaxies compared with the expectations of the standard
hierarchical model for galaxy formation \citep{kly99}\citep{moo99}.  We 
suggested that the deficiency with respect to the standard model might be greater at
lower densities \citep{tul02}, hereafter TSTV.  There might be an expectation 
that there are
many low mass dark matter halos floating about with little or no gas and stars.
We speculated that there might be a possibility to identify the dynamical
manifestations of such halos in a certain special environment.  These low
mass dark halos would not be of sufficient consequence to be noticed in our 
standard groups, even those as modest as the Local Group.  At a mass and
density regime well below that of the Local Group, according to the hypothesis
that we are pursuing that low mass halos are voided of gas and stars, then 
there would be nothing to be seen.  However, in the interval {\it between}
the normal groups and putative much smaller structures it might be expected
that {\it there are groups that are mostly, but not entirely dark.}

As discussed in TSTV, it occurred to us that the `mostly dark'
groups that could be anticipated from theoretical grounds might be found among 
the subset of associations in the T87 study with group-like dimensions and low
velocity dispersions but high implied $M/L_B$ values.  We drew attention to four
such groups thought to lie within 5 Mpc.  

The key issue regarding these `groups of dwarfs' that we identify is whether
they are bound.  {\it If they are bound then they must have high $M/L_B$ values.}
In order to illuminate this issue, we have been involved in programs to 
identify other potential group members and to get good distances to individual
objects.  These galaxies are all near enough that they are easily resolved 
with Hubble Space Telescope (HST) and distances can be established from the
luminosities of stars at the tip of the Red Giant Branch \citep{lee93}, the
`TRGB' distance method.  We first review the present status of these programs.

\section{New Data on Candidate Groups of Dwarfs}

At the time of the paper by TSTV, good distances were only known for members
of one of the four putative groups of dwarfs.  Now there are reasonable 
distances from the TRGB method based on HST images
for {\it most of the galaxies in all of the groups}.  The new distances are
reported in the following publications: \citet{doh98}, \citet{free01},
\citet{kar02c}, \citet{kar03a}, \citet{kar03b}, \cite{mai02}, and \citet{men02}.

The updated results are summarized in Table~2 for the four candidate groups
of dwarfs and one additional comparison group.  No weighting is applied in 
the calculation of the dimension, velocity, and mass parameters since it is 
assumed that the individual galaxies are inconsequential test particles
in the potential of the group.
The following information 
is provided in the table.  (1) Group name from \citet{tul88}.
(2) Prominent galaxy in group.
(3) Number of candidate group members.
(4) Mean group distance.
(5) Inertial radius $R_I^{3D}=(\sum_i^N r_i^2/N)^{1/2}$ where $r_i$ is the 
3-dimensional distance of a galaxy from the group centroid.
(6) Radial velocity dispersion $V_r=(\sum_i^N v_i^2/(N-1))^{1/2}$ where $v_i$
is the radial velocity difference between a galaxy and the group mean 
(7) Blue Luminosity.
(8) ``Projected mass estimate'' 
$M_{pm} = {f_{pm}^{3D} \over G (N-\alpha)} \sum_i^N r_i v_i^2$
\citep{hei85} where $f_{pm}^{3D}=5$ (TSTV) and $\alpha=1.5$.
(9) Virial mass estimate 
$M_v = 3 {(N-1)\over N} V_r^2 R_G / G$ 
where $R_G = N / \sum_{pairs} (1/r_{ij})$ and
$r_{ij}$ is the separation between pairs in the group counted only once.
The factor 3 gives the statistical conversion from the observed radial
velocity to 3-dimensions. Radii are already provided in 3-dimensions because 
of the availability of distances.
(10) Mass to light ratio based on the projected mass estimator.
(11) Mass to light ratio based on the virial analysis.
(12) Mass to light ratio given by TSTV.
(13) Crossing time as a fraction of the Hubble time 
$t_{\rm x} H_0 = 38 R_I^{3D}/V_r$ where $R_I^{3D}$ is in Mpc and $V_r$ is in
km~s$^{-1}$ and $H_0 = 75$ km~s$^{-1}$~Mpc$^{-1}$.

In TSTV, only the projected mass estimator of \citet{hei85} was considered.
Here we also provide the unweighted virial mass estimator.  There can be 
circumstances with small numbers where the virial mass estimator gives
ambiguous results; see the case of the NGC 3109 Group discussed below.
However the use of the two mass estimators provides a better feel for the
considerable uncertainties.

The $M/L_B$ values determined by TSTV are carried for convenience into Table 2.
It can be seen that $M/L_B$ values have increased slightly overall with 
the current analysis.  A principal cause of this small increase was a 
mistake in corrections to luminosities of galaxies in TSTV.  In that paper the 
luminosities had corrections for obscuration based on the prescription by
\citet{dev91}.  However, the obscuration in dwarf galaxies is sufficiently
low as to be unmeasurable \citep{tul98b}.  In this paper, no correction
is made for obscuration within galaxies with $M_B>-16$.  The result is that
luminosities are systematically lowered and $M/L_B$ values are increased. 
Corrections are
made for obscuration due to lines of sight through our Galaxy \citep{sch98}.
In any event, the luminosities of dwarf galaxies are often poorly known.
Errors in the group luminosities of up to 30\% are possible, but errors at
this level do not bring the conclusions of this study into question.
Here are comments on the situation in individual groups.

\noindent{\it 14+12 = NGC 3109 Group:}
The existence of this nearby entity as a distinct group was suggested by
\citet{vdb99}.  In our earlier analysis we entertained that GR8 = DDO 155
might be a member but this galaxy is significantly farther away than was
assumed by TSTV, at 2.24 Mpc rather than 1.51 Mpc \citep{doh98}.  This galaxy
is also, by a substantial amount, the farthest removed from the other 
group candidates on the plane of the sky.  Consequently, we eliminate this
object from consideration as a group member.  Moreover the object 
LSBC D634-03 is now revealed to be a background galaxy coincident with
an HI high velocity cloud.
The other 4 candidates are still
considered at the distances given in TSTV.  In the case of this group,
the virial mass estimator is particularly untrustworthy because two of the
candidates are very close neighbors; NGC 3109 and Antlia are separated by
29 kpc.  Since $R_G \sim 1/\sum (1/r_{ij})$, the single pair dominates 
by an order of magnitude over the next pair in the
calculation of the scale of the potential well.  If the separation of this
pair is arbitrarily set to 100 kpc, still smaller than any other separation
in the group, then the virial mass estimate doubles.  There is a chronic 
ambiguity in the
virial mass estimator with small numbers and instances of close pairs.
However, only this one group suffers this problem among the cases we are
considering.  Overall, the viability of the 14+12 candidate group remains
as it was with the discussion by TSTV, with an implied $M/L_B \sim 400$ if
the entity is bound.

\noindent{\it 14+8 = UGC 8760 Group:}
Distance measurements have revealed that this group is much closer than the
$\sim 5$~Mpc that we expected.  There are now TRGB distances for all 
the candidates, including a new and provisionary one for 
UGC 8760 itself.  Being closer, the linear separations are smaller than we 
appreciated.  Also we
have become aware of a fourth system, UGC 9240, which also is probably
associated.  The three well measured distances are very similar (UGC 8651 at 
3.01 Mpc; UGC 8833 at 3.19 Mpc; UGC 9240 at 2.79 Mpc).  The first two were
associated with the group by TSTV and it is a powerful claim for the reality
of the group that they both have unexpectedly low and similar distances.
A nice test for the hypothesis being pursued is that the distance of UGC 8760
be also low and an HST observation returned as this paper goes to
press reveals UGC~8760 to be at $\sim 3.2$~Mpc.
It is seen in Table~2 that the new $M/L_B$ values
are consistent between methods and considerably larger than found previously.
The difference from the TSTV results is most marked in this case.  Part of this
increase is a direct consequence of the downward revision in distance; 
$M/L_B$ is inversely dependent on distance.  The other causes are the
addition of the 
fourth candidate member and the changes to luminosities discussed above.
A value $M/L_B \sim 1000$ is required if the entity is bound.

\noindent{\it 14+19 = UGC 3974 Group:}
All four candidate members now have TRGB distance measures and they all turn 
out to have distances similar to our expectation of $\sim 5$~Mpc (UGC 3755 at
5.0~Mpc; UGC 3974 at 5.2~Mpc; UGC 4115 at 5.5~Mpc; KK98 65 at 4.5~Mpc).  These
galaxies are rather at the limit of the TRGB method with the HST WFPC detector in
one orbit so the uncertainties are larger in these distances.  Even so,
contributions to the scale of the group from distances and projected
separations are comparable.
Again $M/L_B$ values are larger than found before.  The earlier
result was based on the projected mass estimator with 2-D positions and 
particularly uncertain luminosities.  The current analysis has revised
luminosities and 3-D positions.  Giving consideration to both mass estimators,
the current information suggests $M/L_B \sim 2000$ if the group is bound.

\noindent{\it 17+6 = NGC 784 Group:}
All 4 candidates have TRGB distances now although those for NGC 784 and 
UGC 1281 are unpublished and poor quality.  The agreement is satisfactory
and give a mean distance of 5.0~Mpc, as was initially anticipated.
(UGC 1281 at 4.8~Mpc, NGC 784 at 4.6~Mpc,
KK98 16 at 5.7~Mpc; KK98 17 at 5.0~Mpc).  The factor 3 larger $M/L_B$ than 
previously is attributable to the correction to luminosities
and the distance differentials that reveal this group
is extended more in the line-of-sight than in projection.  
If the group is bound then $M/L_B \sim 1000$.

\noindent{\it 14+13 = Foreground Sculptor Group, a comparison case:}
This structure was included in the discussion by TSTV.  In the past, several 
of the galaxies have been considered as part of the Sculptor Group
but we consider this historical group to consist of two distinct entities.
We give attention to the nearer part.  One good new
candidate is added to the group specified by TSTV: ESO 294-010.  All but 
the group's most luminous galaxy, NGC 55,
have measured distances now (ESO 294-010 at 1.92~Mpc; NGC 300 at 2.00~Mpc;
UGCA~438 at 2.23~Mpc; IC~5152 at 2.07~Mpc).  Results are little different
than previously.  The mass estimates are as low as for any of the groups
discussed above because of the very low velocity dispersion.  However there
are two moderate galaxies in the group so there is substantial luminosity.
IC~5152 is removed by 770~kpc from the group centroid, mostly in projection,
so the status of this object is in doubt.  Whether or not it is included, if
the entity is assumed to be bound then $M/L_B \sim 17$.

\section{Comparison of Dwarf Group Properties}

The candidate groups have mass estimates in the modest range 
$1-9 \times 10^{11} M_{\odot}$.  Characteristic dimensions range 
290~kpc~$ < R_I <~$ 540~kpc with a mean of 420~kpc, typical of more familiar
groups (T87).  Velocity dispersions range from 36~km~s$^{-1}$ down to 
an incredibly
low 13~km~s$^{-1}$ with a mean of 22~km~s$^{-1}$.  These values are all 
very low compared to familiar luminous groups and account for the low
mass estimates.  Luminosities are all in the decade 
$1-10 \times 10^8 L_{\odot}$ except for the comparison case of the 14+13
or Foreground Sculptor Group which has almost an order of magnitude more
luminosity than the most luminous of the others.

Crossing times as a fraction of the Hubble time are given in the last column
of Table~2.  Typically values are $\sim 0.8 {\rm H}_0$.  Of course,
the crossing times are long because the velocity dispersions are so low.
They are a sufficiently large fraction of the age of the Universe that 
one can worry that velocities are simply attributable to the expansion of
the Universe; ie, that these are not bound groups.  Only in the case of
the tight 17+6 (NGC 784) Group is the crossing time substantially less than
a Hubble time, H$_0^{-1}$.  If IC~5152 is accepted as part of the 14+13 
(Foreground Sculptor) Group then the crossing time for this group is 50\% 
longer than H$_0^{-1}$, but this is the comparison case: a group with low 
$M/L_B$.

To this point there has been almost no discussion of uncertainties.  First,
it is to be emphasized that the measurement errors in the observed parameters
are of almost no consequence.  The radial velocities are obtained by 
observations of the 21cm line of HI and for these dwarfs are determined with
an accuracy of $\pm 5$~km~s$^{-1}$.  The distances are determined with the
TRGB method.  There might be a distance scale zero point issue at the level of 
10\% that would shift the entire sample in a similar fashion and not affect
this discussion.  Relatively, the distances should be accurate to 5\% for
galaxies within 3 Mpc.  There is a degradation to $10-15\%$ at 5 Mpc.
However in none of the groups discussed here is the depth of the group
dominant over the projected dimensions.  Hence errors in measured
dimensions do not make a significant contribution to uncertainties.  The most
poorly known direct observables are the luminosities.  The global group
luminosities attributable to the identified candidates can have errors as
large as 30\%.  The luminosities come from heterogeneous sources, some with
large errors.  The situation could be improved with dedicated 
observations.  Still, the real uncertainties lie elsewhere.

The greatest uncertainty of all is whether or not the groups are bound.
Assuming for the moment that they are, then there are four dominant 
uncertainties in the calculation of a group mass: (a) small number statistics,
(b) the statistics of velocity deprojection from one to three dimensions,
(c) potentially poor coverage of the group gravitational well, 
and (d) the nature of the galaxy orbits.  The first two of these problems
lead to uncertainties that are statistical in nature.
The groups under consideration
have 4 or 5 candidate members each.  The number of pairs grows as 
${N(N-1)/2}$
so each additional candidate gives a big improvement.  As for the velocity
deprojection problem, recall that mass calculation involves $V^2$ and the
deprojection correction is a factor 3.  The latter two problems raise 
possibilities of systematic errors in the mass calculations.
With regard to the question of
coverage of the group potential well, one aspect is simply the limited 
coverage with only 4 or 5 test particles and another aspect is the uncertainty
in the limits of the putative bounded regions.  Do the candidates at larger
radii represent those limits, or are they frequently beyond those limits and
escaping?  Finally, there is the nature of the galaxy orbits.  Are they
isotropized to a degree as would occur as a system approaches virialization
or are they largely radial as would be expected of a group still in formation,
with orbits dominated by infall?  The alternatives have different implications
for the projected velocities.  
Following the assumption of the virial theorem: $2T/|W| = 1$, where the
kinetic energy per unit mass is $T = {3(N-1) \over 2N} V_r^2$ and the potential 
energy per unit mass is $W = -G M_v / R_G$ (see section 6).  The
possibility of systematic departures from the virial condition was
discussed in section 3.
Both from 
analytic and N-body simulations \citep{mer03} it is found that mean values
of $2T/|W|$ evolve from small values at turnaround to $\sim 1$ at first collapse,
to ratios up to 20\% above unity, whence there is a slow approach to
virialization from values above 1.  There can be large excursions in individual
cases.  The results of this paper could require modification due to this probable
systematic effect.  Dense groups that are well advanced toward virialization
will tend to have $2T/|W|$ greater than unity by $0-20\%$ so mass estimates
will be above true values while groups that are still collapsing will tend
to have $2T/|W|$ less than unity by up to a factor 2 so mass estimates
will be below true values.

Given the other large sources of errors, the uncertainty regarding the 
dynamical state is not dominant so the virial condition will be assumed.
Mass estimates have been calculated two ways, following the discussion by
\citet{hei85}.  Except in the special case of the 14+12 (NGC 3109) Group
that received discussion, the two mass estimates for a group differ from the 
mean by no more than 35\%.  If it is simply assumed that the group is bound
then one gets a {\it lower mass limit} which is one-half the virial estimate.
In the plots that will be shown further along, 
factor 3 uncertainties in mass have been assumed for groups with 4 candidates.


Even the very large uncertainties that have been discussed do not bring into
question the fundamental claim being made pertaining to the low density regime.
{\it If the candidate dwarf groups are bound then high $M/L_B$ values are 
required for the groups.}  We can think of two arguments that favor
the point of view that 
the groups are bound.  The first notes the continuity in properties
with well established groups in terms of dimensions and velocity dispersions.
These properties are illustrated in Figure~5.  To a reasonable degree,
the E/S0 dominated groups and spiral dominated groups separate on
this plot.  The early-type groups tend to be restricted spatially but
can manifest large velocity dispersions while the late-type groups
inevitably have modest velocity dispersions but can be dispersed
spatially (there are a few groups with majority early types with the
properties of the spiral groups).
The candidate dwarf groups can be viewed
as an extension of the distribution of the groups dominated by spirals.
There is nothing to suggest from these
parameters that characterize the dynamical state that the dwarf groups are
other than a continuation of the family of bound groups to lower masses.
It is only when attention is given to the luminosities of the systems that
one appreciates that the situation is quite different from the familiar.

The second argument draws on the highly correlated nature of the distribution
of dwarf galaxies.  Within 5 Mpc, most known dwarfs are members of the few
well established groups (14-7: CVn~I Group; 14-10: M81 Group;
14-11: Maffei/IC~342 Group; 14-12: Local Group; 14-13: Sculptor Group; 14-15: 
Centaurus Group) and most of the rest are associated with a few relatively
isolated bright galaxies or are members of the dwarf groups that
have been identified.  The current census of the sky for dwarf galaxies
is mainly due to the hard work extracting candidates from the Second Palomar
Observatory Sky Survey in the northern sky and the ESO/SERC survey of the
southern sky by Karachentseva and collaborators
\citep{kaa98}, \citep{kaa99}, \citep{kar00}, \citep{kaa00}, with HI 
follow up reported in \citet{huc01} and earlier papers.  The HI Parkes
All Sky Survey (HIPASS) is providing an independent search of the southern
part of the sky \citep{bar01}.

\realfigure{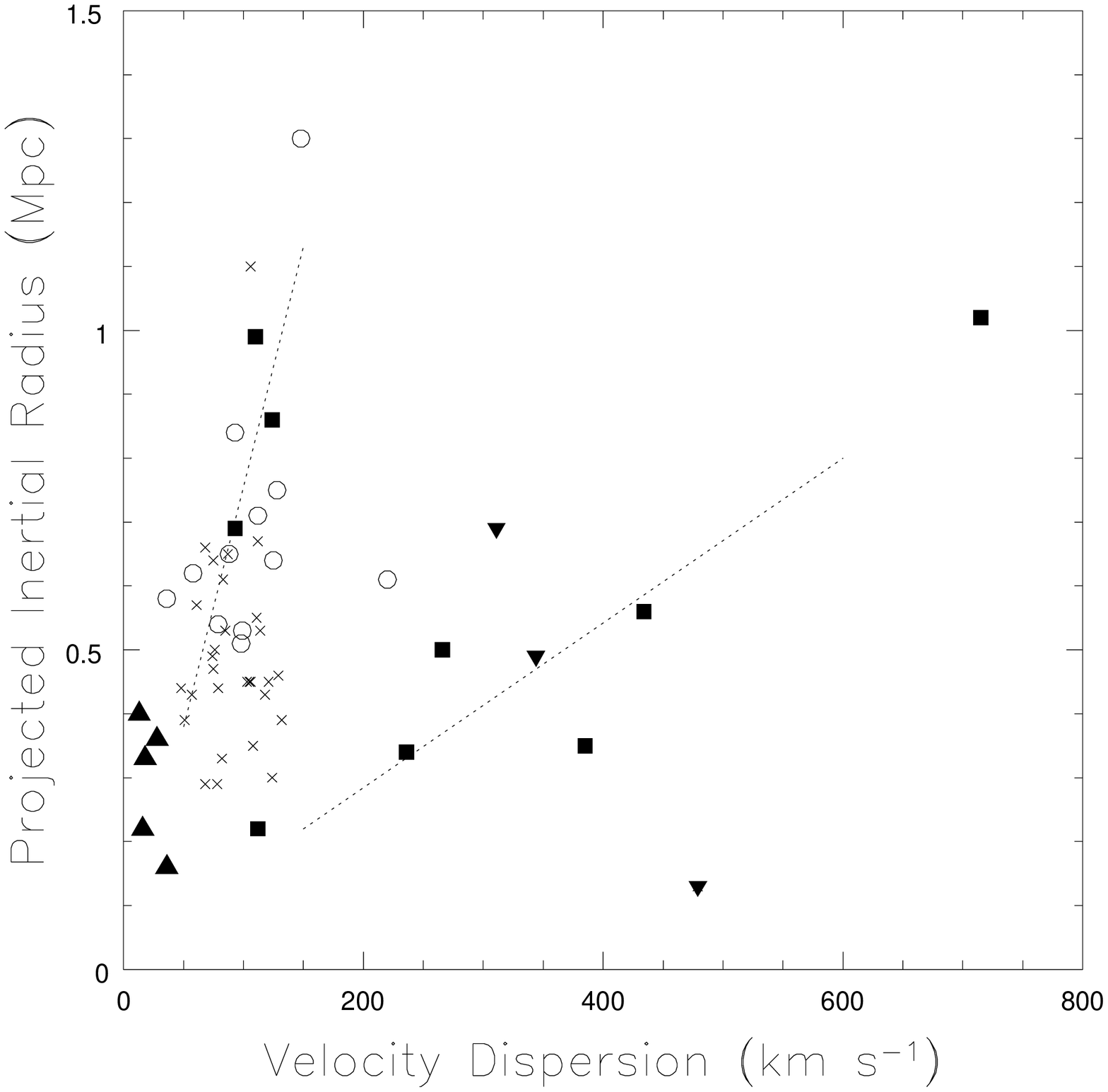}{Group dimensions compared with group velocity dispersions.  Groups
dominated by early type galaxies: filled 
squares and inverted triangles. Groups dominated by late types: open circles.
Small groups: crosses.  Groups of dwarf galaxies (and 14+13 group):
triangles.
The separate trends of the early and late type groups are indicated by
the dashed lines.}{fig5}


\citet{kar03a} provide a tabulation of 156 galaxies, big and small, identified 
to exist between 1 and 5.5 Mpc of us (hence excludes the Local Group).
Fully 101 of the objects are associated with the well established groups
identified in the previous paragraph.  Another 13 are associated with
relatively isolated luminous galaxies and 6 are associated with somewhat more
distant groups or are at very low galactic latitude.  Then there are the 21
galaxies we associate with the groups of dwarfs or the comparison 14+13
(NGC 55) Group.  That leaves only 15 low luminosity galaxies remaining
through the rest of the volume.
Several of these are in close proximity to one another.

The strong correlation in position of the nearby galaxies is illustrated 
in Figure~6.  The top panel shows the two-point correlation function for
all known galaxies at $\vert b \vert > 28$ within 5 Mpc, excluding the Local 
Group within 1.1 Mpc.  Good distances exist for a majority of these galaxies,
and rough distances exist for the rest, so what is shown here is the 
three-dimensional two-point function.  Normalization was achieved by 
comparison with 1000 monte carlo random distributions within the volume.
The strong positive correlation at separations less than
1 Mpc is dominated by the contributions from the well known nearby 
galaxy groups.

\realfigure{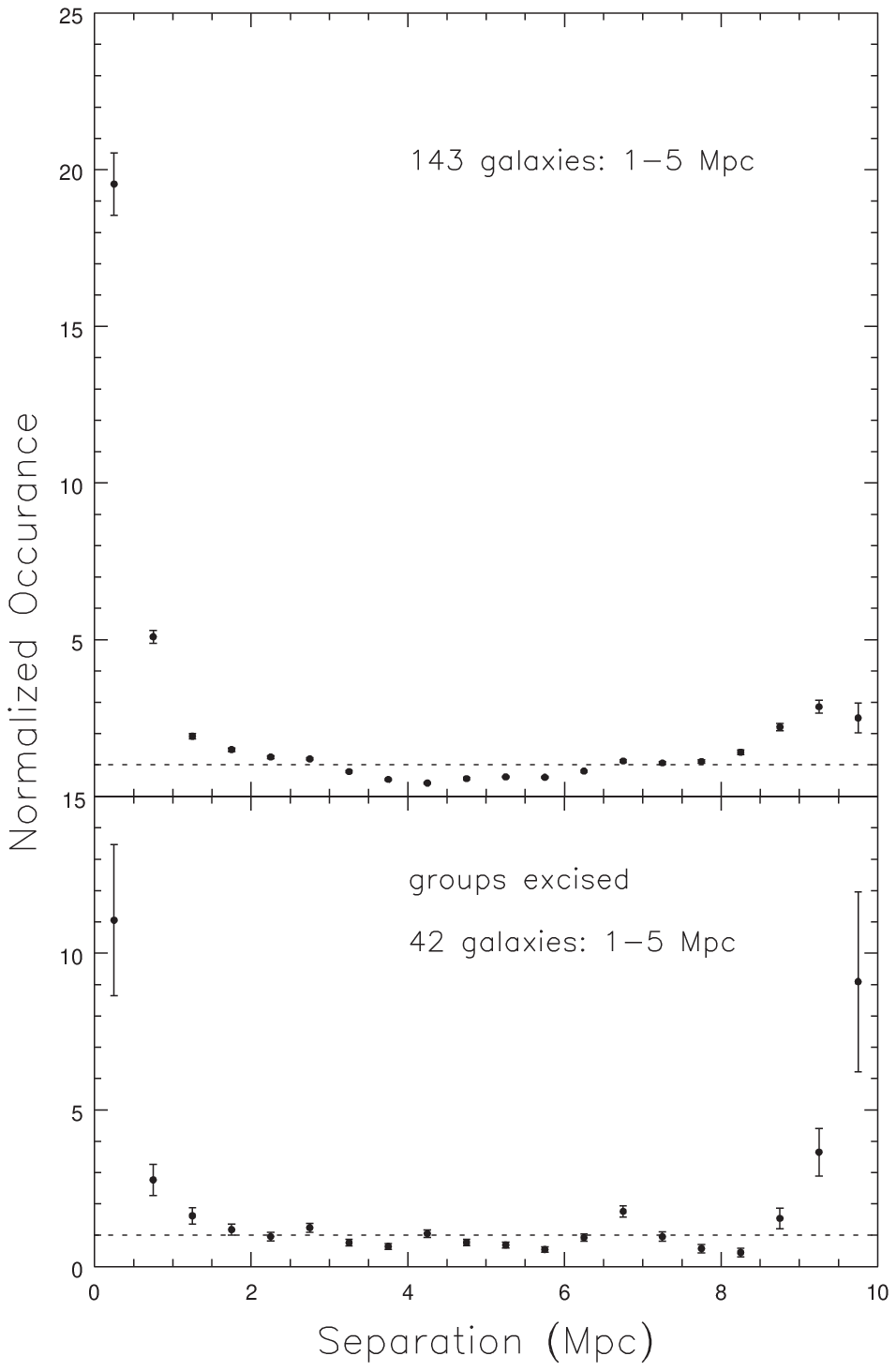}{Two-point 3-dimension correlation functions.  Top panel: correlation involving all 143
known galaxies within $1.1 < d < 5$ Mpc and $\vert b \vert > 28$.
Bottom panel: correlation with 42 galaxies after excising volumes containing
five big groups.}{fig6}


The bottom panel in Fig. 6 examines the two-point correlation when 
contributions from the
dominant groups have been excised.  In this case, spheres of radius 1.4 Mpc
have been defined centered on the 5 dominant nearby high latitude groups:
those associated with M81, CVn~I, Centaurus, Sculptor, and NGC 1313 
(the Maffei Group lies at low latitude as
does a majority of the Centaurus Group).  The 80\% of the volume that 
remains (220 Mpc$^3$) contains 30\% of the original sample (42 galaxies).
The Foreground Sculpter and three of four of our candidate dwarf groups 
are in the remaining volume, contributing 17 of the 42 galaxies (the 14+19
Group lies at low Galactic latitude; in fact the fussy $\vert b \vert > 28$
limit was chosen to allow inclusion of the entire 17+6 Group). 

It is seen that even when the historically known groups are eliminated
there is still a strong correlation signal at separations less than 1 Mpc. 
The signal at separations less than
0.5 Mpc is 63\% of the signal in the top panel.  This signal is dominated
by the correlations within the four low mass candidate groups.

The positive correlation signal at separations greater than 9 Mpc has an
easy explanation.  The monte carlo normalization assumes objects are 
distributed randomly throughout the available volume but in fact most nearby
galaxies lie in the supergalactic equatorial plane.  The signal at large
separations comes from cross-correlations from opposite ends of the 
equatorial plane, between the 17+6 members and
galaxies in the region of the Centaurus Group on the one hand and
between galaxies in the vicinity of the Sculptor Group with those near
CVn~I and near Centaurus.  In this small sample, only 3-4 galaxies in each 
of these regions is enough to create the spurious cross-correlation signal.
The weaker signal at large separations in the top panel arises from the 
same source: the concentration of local galaxies to a plane. 

The present analysis is preliminary because 
the census of dwarfs is still
uncertain.  It is still not clear that even the high latitude sky has been
uniformly surveyed.  It is not clear how rapidly candidates are lost with 
Galactic latitude.  HI signals for objects with velocities near zero can 
be lost in the confusion of Galactic emission.
Our knowledge of the distances
of candidates is improving rapidly with HST imaging but is still very
incomplete.  On the basis of present information the dwarf groups that have
been identified are manifest enhancements over a random distribution.
They correlate comparably well as the galaxies in previously established groups.

\section{Everything Together}

The small dimensions of the dwarf groups and the highly significant
2-point correlation signal on scales $<1$ Mpc suggest that the groups of 
dwarfs are bound.  Given that 
proposition, we can add the 
information from the low density regime to the data from the intermediate
and high density regime shown in Fig.~2.  The combined data is seen in 
Figure 7.  T87 groups of 2-4 members are added to provide information
about the situation at low luminosities, though mass estimates are very
uncertain for these entities. The four groups of dwarfs are plotted with
error bars in mass.  They lie at luminosities below $10^9 L_{\odot}$
and well below the dotted $M/L_B$ constant line.  The comparison 14+13 (NGC 55)
Group lies in the same mass range but at higher luminosity, well above the
dotted line.  The solid curve was fit to the ensemble of data.  The curve
corresponds to the expression:
\begin{equation}
L_B = \phi M^{\gamma} e^{-M^{\dag}/M}
\end{equation}
This equation relates luminosity and mass with 3 constraints: a logarithmic 
slope at the
high mass end, $\gamma$, a low mass exponential cutoff set by $M^{\dag}$,
and a normalization, $\phi$.  The curve was fit by minimizing a $\chi^2$ with
uncertainties taken in mass only:
\begin{equation}
\chi^2 = \sum_i^N ({{\rm log}M_i - {\rm log}M_{fit} \over {\rm log}\sigma})^2
\end{equation}
Here, $M_{fit}$ is the mass in the relationship described by eq.~(2)
at the luminosity $L_i$ of group $i$, with measured mass $M_i$.  
The $\chi^2$ normalization is given the dependence on the number of group
members $N$ according to the formulation ${\rm log}\sigma = A/N^{0.75}$.  
A weighting that favors
groups with large $N$ more than the statistical $N^{-0.5}$ is justified by
the expected better approximation to a virialized state for the larger
groups and to compensate for the small numbers of large $N$ systems.
The absolute normalization of the $\chi^2$ evaluator is arbitrary: we 
simply search for the lowest value.  We set $A=2$ for the traditional
groups and $A=1$ for the groups of dwarfs; ie, give double weight to the
groups of dwarfs because they are drawn from a restricted region and are 
few in number.

\realfigure{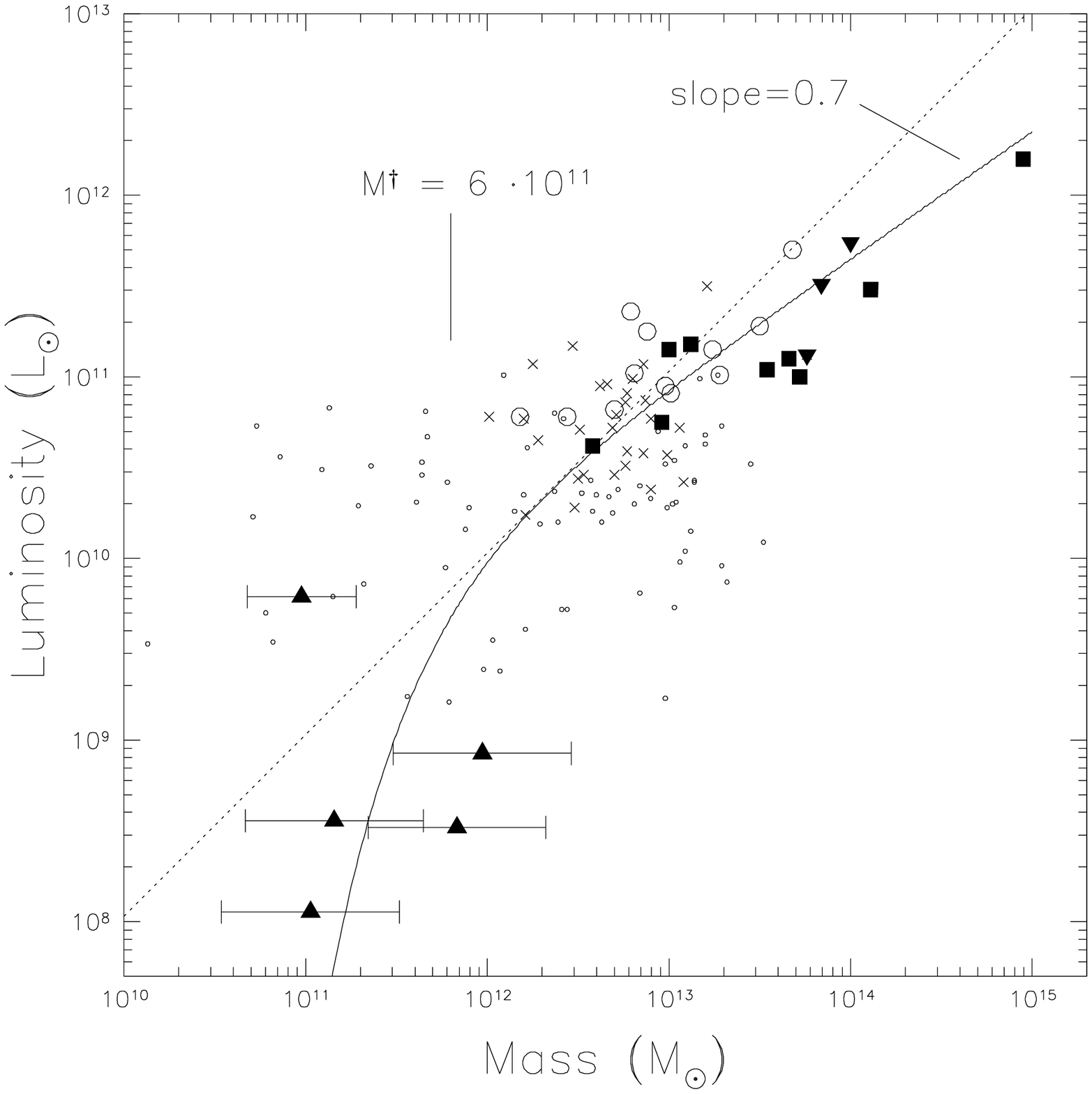}{Mass vs. light for groups over the full range of density regimes.  The data
for the higher density regimes are the same as seen on the right panel of
Fig.~2 with the addition of the small open circles which represent groups
with 2--4 identified members. 
The 5 groups that explore the low density regime are plotted with
error bars.  The $M/L_B=94$ dotted line is carried over from Fig.~2.  The solid
line is the fit described in the text with a high mass slope of $\gamma=0.7$
and an exponential fall off at masses less than 
$M^{\dag} = 6 \times 10^{11} M_{\odot}$.}{fig7}


The curve superimposed on the data in Fig.~7 has a slope at the
high mass end of $\gamma = 0.70$.  It can be appreciated that taking errors
only in mass leads
to a steeper slope than if errors are distributed into luminosity.
At the low mass end the curve demonstrates an exponential cutoff 
characterized by $M^{\dag} = 6 \times 10^{11} M_{\odot}$.  
The normal luminous groups with 5 or more members only provide
information at masses $>10^{12} M_{\odot}$.  Groups of 2--4 members provide
information at lower luminosities and masses but at the cost of large
uncertainties in mass.  The evidence for the cutoff comes from the four
candidate dwarf groups.
In the absence of the dwarf groups we might be
left to suppose that there is simply a lower limit to the mass range of 
groups.  However, from Fig.~7 we instead infer that smaller mass 
groups do exist but groups are not manifested by the 
light of stars at very low masses.
The break is more dramatically represented in Figure 8.  This plot shows 
the same
data but the vertical axis gives the departure from a $45^{\circ}$ line in
the previous plot.  Large errors in masses in small groups create diagonal
scatter in this plot.  The inference we draw from the Foreground Sculptor
Group, with very low $M_B/L$, is that there is tremendous real scatter
in the baryon content of halos in the proximity of the mass cutoff.

Interestingly, both \citet{mar02} and \citet{vdb03} have deduced a similar 
dependence between 
mass and light from observations of the luminosity function of galaxies
derived from redshift surveys and the assumption that the underlying halo
mass function follows the modified Press-Schechter description of \citet{she99}.
The deduction follows from the observation that, compared with the mass
function, the luminosity function
is shallower at the faint end and cuts off more abruptly at the bright end,
implying $M/L$ increases at the two extremes relative
to the intermediate range.

\realfigure{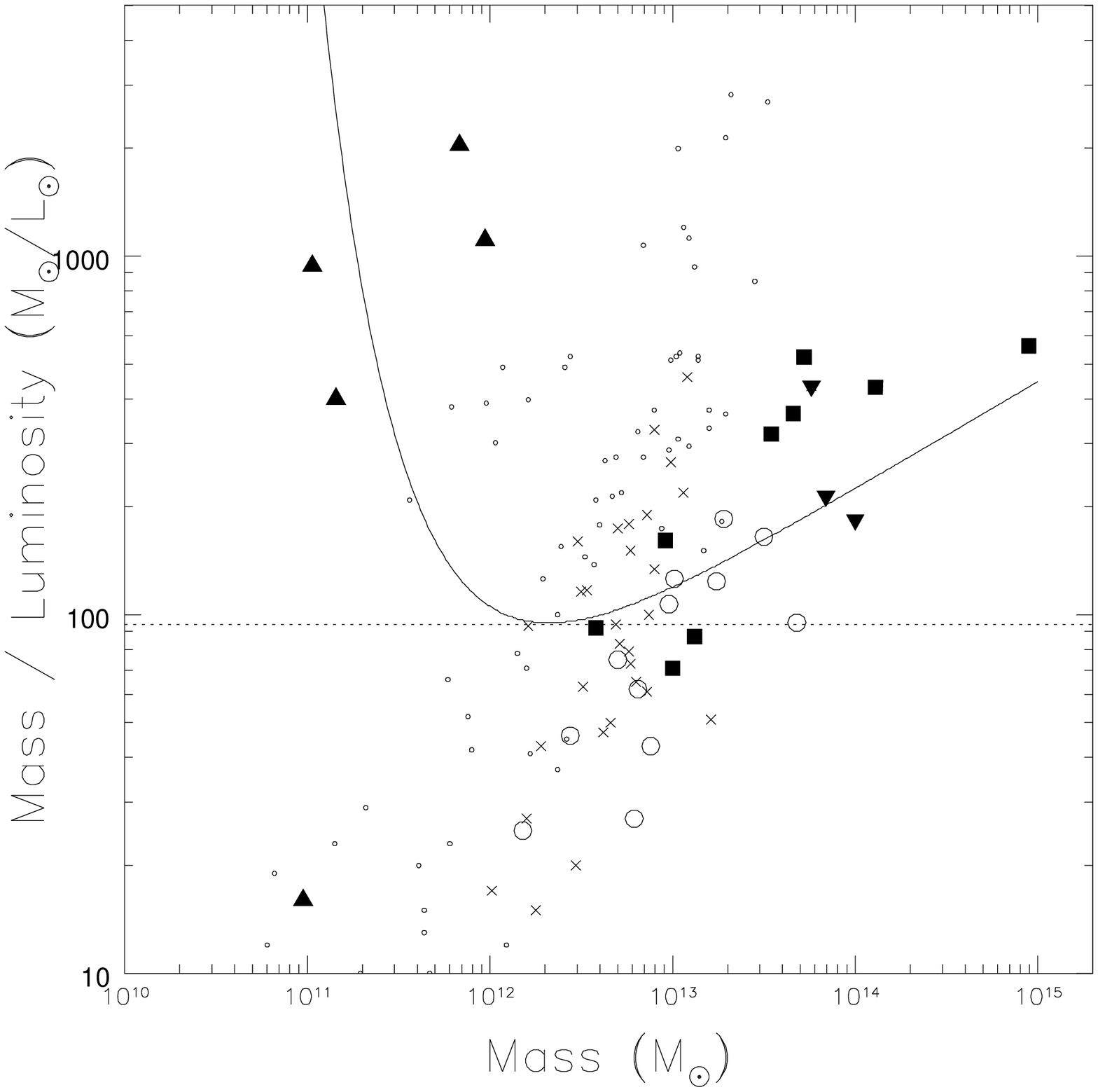}{Mass to light ratio vs. mass over the full range of density regimes.  The data
are the same as in Fig.~7.
The solid curve and dotted line are transpositions of the curve and line in 
Fig.~7. }{fig8}


To conclude, we take a look at the mass function in the volume with
$\vert b \vert >30$ and 
a distance limit of
$25 h_{75}^{-1}$~Mpc (distances from a numerical action velocity field
model)
and the fraction of the total mass lying in specific environments.  
This information is summarized in Figure 9.
The mass function incorporates all the groups involved in the earlier
discussion, plus contributions from galaxies that lie outside any of these
groups.  We have no information on the extended mass around individual galaxies
but for the purpose of constructing this mass function it is assumed that
$M/L_B^{single}=100 M_{\odot}/L_{\odot}$.  The contribution of the individual 
galaxies
by number is given by the figures in brackets in Fig.~9.  It is seen that 
the individual galaxies only become dominant in their contribution below
$10^{12} M_{\odot}$ and with our $M/L_B$ assumption for the individual galaxies
their contribution to the overall mass budget is small.  

The mass function shown in Fig.~9 becomes incomplete below $10^{12} M_{\odot}$.
The error bars only reflect the statistical uncertainty associated with the
observed contributions.  To get an estimate of the contribution to the total
mass budget of low mass groups we consider the restricted volume within
5~Mpc.  In this volume there are 5 groups of luminous galaxies, excluding the
low latitude 14-11 (Maffei/IC~342) Group, and these groups cumulatively
contain about $12 \times 10^{12} M_{\odot}$.  Three high latitude luminous 
galaxies unassociated with significant groups (NGC 1313, NGC 3621, and 
NGC 6503), if assumed to have an associated $M/L_B=100$, would contain
about $2 \times 10^{12} M_{\odot}$ in their vicinity.  The five low mass
groups discussed here would cumulatively contain about 
$2.0 \times 10^{12} M_{\odot}$.  Roughly $3/4$ of the mass related to
observable galaxies within 5 Mpc would lie in the major groups, almost
90\% would be associated in some way with luminous galaxies, and only about
12\% would be in the regions of the groups of dwarfs.  Although the
number of groups 
associated with low mass halos are comparable to the number of
familiar groups in the 
local region, they contribute only a small fraction of the mass.  Hence the
inventory of the mass of the Universe tied up in bound structures represented
by Fig.~9 is unlikely to be seriously in error at the low mass end.

\realfigure{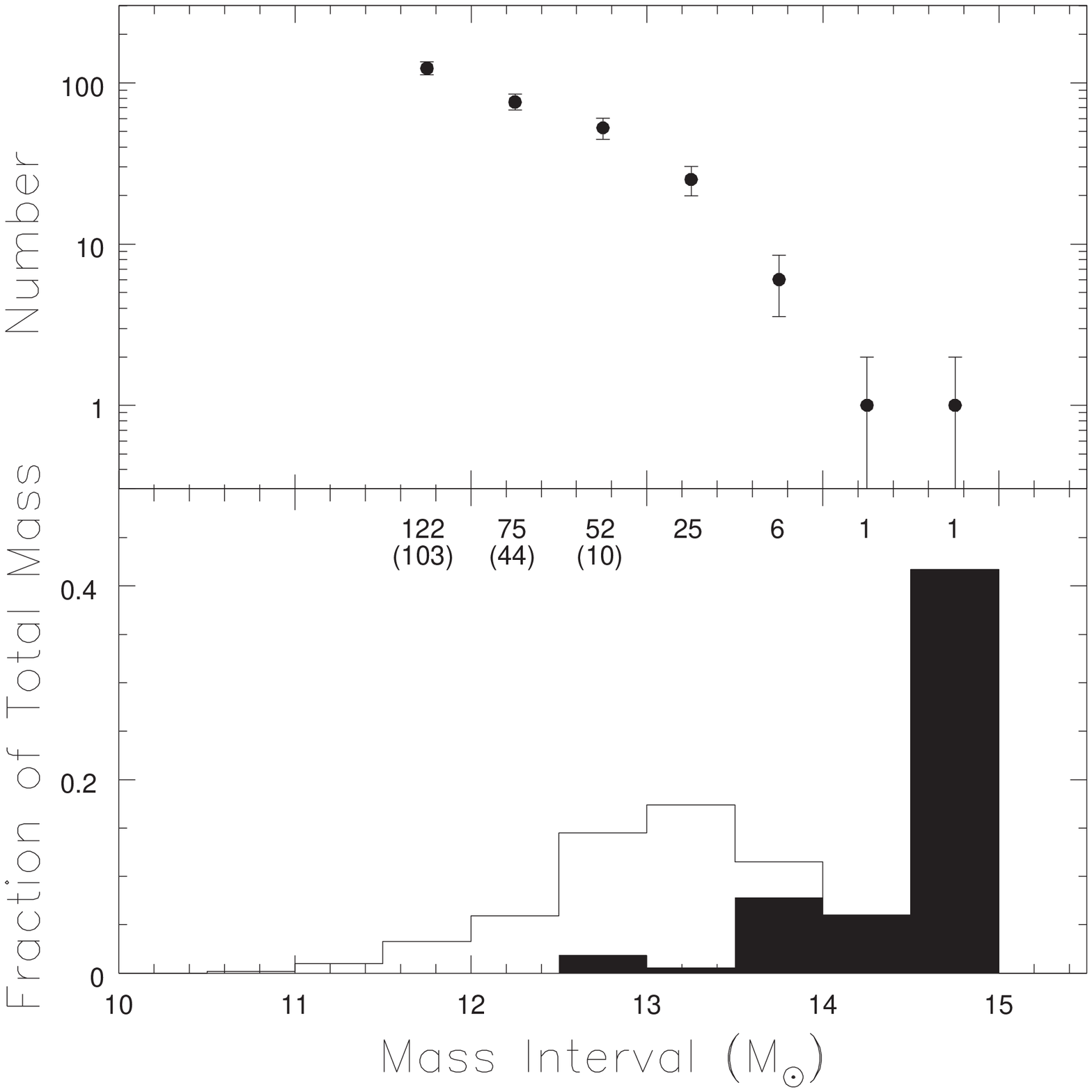}{{\it Top panel:} Observed mass function in the volume within 
$25 h_{75}^{-1}$~Mpc 
and $\vert b \vert > 30$.  The numbers of objects in each bin are given along the
top of the lower panel.  In brackets are the numbers of individual
galaxies; ie, galaxies outside of groups.  {\it Lower panel:} Fraction of the
total mass associated with the mass function in half-dec bins.  The mass
associated with the groups of predominantly early types is indicated by the
filled histograms.  The open histograms correspond to the mass found in the
groups with predominantly late types.}{fig9}


The histogram of mass fractions by mass interval seen in Fig.~9 can be 
summarized as follows.  Fully 40\% of the mass in this local volume is in the
single object, the Virgo Cluster (clearly, the high mass end of the mass
function is poorly determined in such a restricted volume).  The groups that 
have short crossing times
and are dominated by E/S0/Sa galaxies (including Virgo) contain 60\% of the
mass.  Looking at groups of all types, 90\% of the mass is in groups with
${\rm log} M > 12.5$.  The luminous groups within 5 Mpc all lie essentially
within the mass bin $12.0 < {\rm log} M < 12.5$, on the tail of the
histogram seen in Fig.~9.  Yet even with respect to such groups, the low 
mass groups within 5 Mpc make only a minor contribution to the mass budget.
It can be concluded that
they do not contain an important fraction of the mass of the Universe.

\section{Summary}

There is surprisingly strong evidence for variations in the relationship
between blue light and mass as a function of environment.  On the scale of
groups, the lowest $M/L_B$ values are found on mass scales of 
$10^{12} - 10^{13} M_{\odot}$ where typical values are 
$M/L_B \sim 90 M_{\odot}/L_{\odot}$.  Groups in this mass range are almost 
always composed of late type galaxies with ongoing star formation.
Groups with higher mass produce less blue light.  The trend is enhanced by
the local density as identified by the crossing time.  High density regions,
those with short crossing times, are darker.  The morphology of the group
members is highly correlated with these trends: groups with
predominantly E/S0/Sa systems have 
short crossing times and large $M/L_B$.  This pattern is consistent with a
picture in which dense regions formed earlier and today star formation is
largely exhausted in these places.  The stellar populations are fainter and
redder.  Possibly in these environments there have been multiple collisions
between galaxies that have scattered many stars into the intracluster 
environment where their
light goes undetected in our light inventory.  Then it is certainly known
that the dense environments of E/S0 groups and clusters glow with
thermal X-ray emission, to the degree that most of the baryons in these
environments are in the hot intracluster gas and not in stars.  Together, these
astrophysical processes could explain the observed factor of 7 increase in 
$M/L_B$ in 
proceeding from spiral dominated groups in the mass range 
$10^{12} - 10^{13} M_{\odot}$ to E/S0 dominated groups in the mass range
$10^{14} - 10^{15} M_{\odot}$.

At the other end of the mass spectrum, below $10^{12} M_{\odot}$, there is
a cutoff in the visible manifestations of groups.  One unlikely possibility
is that there simply are few groups with less mass than $10^{12} M_{\odot}$.
A more likely possibility is that group halos of lower mass exist but are
difficult to identify because of a deficiency of light.  We  have presented
evidence that such low mass groups might be common.  The number of such
groups that can be identified within 5~Mpc are comparable to the number of
luminous groups.  The deficiency of light is very great in these cases,
assuming that the groups are bound.  Groups in the mass range 
$10^{11} - 10^{12} M_{\odot}$ can have $M/L_B$ values 5 to 20 times higher than
the groups of spiral galaxies in the range $10^{12} - 10^{13} M_{\odot}$.
The possibility is open that there may be halos in the dwarf group mass range
and lower that are totally invisible.

As an aside, if the hypothesis of this paper is correct that groups of dwarfs
are bound then the two decade old controversy between the alternatives of 
dark matter versus non-Newtonian gravity (Milgrom 1983) is definitively 
resolved.  It is implausible that the groups of dwarfs are bound by stars
or gas.  The relationship between the observable constituents and the 
gravitational field is quite unlike that of more luminous groups. 

The transform between light and mass is being shown to be complex.
Of particular cosmological importance is the sharp trend toward darkness at
high mass.  The Virgo Cluster has 40\% of the mass attributed to groups in
the volume of our study but it only contributes 15\% of the blue light.
The rich clusters have a dynamic importance much greater than would be
expected from their light.  By contrast, though there is evidence that
low mass groups may be rendered almost or completely invisible, their
cumulative contribution to the clustered mass of the Universe seems to be
small.  To the limit that it can be traced, the mass function is too flat
for low mass structures to make a substantial contribution to the 
inventory of mass.

\acknowledgments

The work over the years involving the Numerical Action modeling has
involved a happy collaboration with Jim Peebles and, especially, Ed Shaya.
My principal collaborator on work at the faint end of the galaxy luminosity
function is Neil Trentham.  Jose Pacheco has provided valuable insight
regarding the dynamic state of unrelaxed structures.
This research is being supported by JPL Contract 1243647 and STScI awards
HST-GO-09162 and HST-GO-10210.

\clearpage



\clearpage
\tablecaption{Properties of Groups from T87. \label{tbl-1}}
\begin{deluxetable}{lcccccccccc}

\tabletypesize{\scriptsize}
\tablewidth{0pt}
\tablehead{
\colhead{Group} & \colhead{$N_{gal}$} & \colhead{\% Early} & \colhead{Dist.} &
\colhead{$V_g$} &
\colhead{$V_r$}  & \colhead{$R_I$} & \colhead{$t_{\rm x} {\rm H}_0$} &
\colhead{log$L_B$}     & \colhead{log$M_{\rm v}$}  &
\colhead{$M_{\rm v}/L$} \\
 &  &  & \colhead{(Mpc)} & \colhead{(km s$^{-1}$)} &
\colhead{(km s$^{-1}$)} & \colhead{(Mpc)} &  &
\colhead{($L_{\odot}$)} & \colhead{($M_{\odot}$)} &
\colhead{($M_{\odot}/L_{\odot}$)}
}
\startdata
11-01 & 174 &  61 & 16.8 & 1042 & 715 & 1.02 & 0.08 & 12.20 & 14.95 &  562 \\
11-04 & ~~7 &     & 13.8 & 1576 & 124 & 0.30 & 0.13 & 10.59 & 12.77 &  151 \\
11-10 & ~~9 &  17 & 23.9 & 1221 & ~79 & 0.54 & 0.36 & 11.36 & 12.79 &  ~27 \\
11-14 & ~~5 &     & 19.2 & ~838 & 106 & 1.10 & 0.55 & 11.50 & 13.21 &  ~51 \\
12-01 & ~57 &  22 & 17.2 & ~967 & 148 & 1.30 & 0.47 & 11.70 & 13.68 &  ~95 \\
12-03 & ~~9 &  22 & 22.9 & 1352 & 112 & 0.71 & 0.34 & 11.01 & 13.28 &  185 \\
12-05 & ~~6 &     & 23.0 & 1384 & ~83 & 0.61 & 0.39 & 10.95 & 12.62 &  ~47 \\
12-06 & ~~9 &  29 & 20.1 & 1020 & 125 & 0.64 & 0.27 & 10.91 & 13.01 &  126 \\
14-01 & ~25 &  52 & ~9.7 & ~911 & 266 & 0.50 & 0.10 & 11.00 & 13.72 &  523 \\
14-04 & ~22 &  14 & ~7.6 & ~596 & ~58 & 0.62 & 0.56 & 10.78 & 12.44 &  ~46 \\
14-05 & ~~8 &     & ~7.3 & ~571 & 129 & 0.46 & 0.19 & 10.77 & 12.90 &  134 \\
14-06 & ~~5 &     & ~7.8 & ~698 & ~78 & 0.29 & 0.20 & 10.44 & 12.50 &  116 \\
14-07 & ~22 &     & ~3.5 & ~309 & ~51 & 0.39 & 0.41 & 10.24 & 12.21 &  ~93 \\
14-09 & ~~9 &     & ~5.0 & ~367 & ~82 & 0.33 & 0.21 & 10.46 & 12.53 &  117 \\
14-10 & ~12 &     & ~3.3 & ~242 & 108 & 0.35 & 0.17 & 10.46 & 12.70 &  174 \\
14-11 & ~~8 &     & ~3.0 & ~188 & ~75 & 0.47 & 0.33 & 11.17 & 12.47 &  ~20 \\
14-12 & ~10 &     & ~0.0 & ~-18 & ~57 & 0.43 & 0.40 & 10.78 & 12.01 &  ~17 \\
14-13 & ~11 &     & ~2.1 & ~197 & 118 & 0.43 & 0.19 & 10.38 & 12.90 &  ~28 \\
14-15 & ~12 &     & ~4.3 & ~304 & ~68 & 0.66 & 0.52 & 10.96 & 12.66 &  ~50 \\
15-01 & ~~9 &  88 & ~7.2 & ~626 & 112 & 0.22 & 0.10 & 10.62 & 12.58 &  ~92 \\
17-01 & ~13 &  ~9 & 10.0 & ~752 & ~36 & 0.58 & 0.85 & 10.78 & 12.18 &  ~25 \\
17-04 & ~~6 &     & ~9.8 & ~813 & 114 & 0.53 & 0.25 & 10.57 & 12.99 &  266 \\
19-01 & ~~8 &     & 10.6 & ~735 & ~87 & 0.65 & 0.39 & 10.86 & 12.76 &  ~79 \\
21-01 & ~11 &  36 & 22.9 & 1087 & 220 & 0.61 & 0.15 & 11.28 & 13.50 &  165 \\
21-03 & ~10 &  ~0 & 20.9 & 1009 & ~98 & 0.51 & 0.28 & 10.82 & 12.70 &  ~75 \\
21-06 & ~12 &  55 & 22.3 & 1207 & 124 & 0.86 & 0.37 & 11.15 & 13.00 &  ~71 \\
21-10 & ~~5 &     & 20.1 & 1082 & ~61 & 0.57 & 0.49 & 10.77 & 12.20 &  ~27 \\
21-12 & ~10 &  30 & 24.5 & 1453 & ~88 & 0.65 & 0.39 & 11.02 & 12.81 &  ~62 \\
41-07 & ~~5 &     & 24.0 & 1219 & 132 & 0.39 & 0.16 & 11.07 & 12.86 &  ~61 \\
43-01 & ~13 &  11 & 19.2 & ~970 & 128 & 0.75 & 0.31 & 11.15 & 13.24 &  124 \\
44-01 & ~~7 &     & 15.9 & ~922 & ~76 & 0.50 & 0.35 & 10.79 & 12.71 &  ~83 \\
51-01 & ~31 &  90 & 16.9 & 1344 & 434 & 0.56 & 0.07 & 11.48 & 14.11 &  431 \\
51-04 & ~17 &  62 & 17.9 & 1427 & 110 & 0.99 & 0.48 & 11.18 & 13.12 &  ~87 \\
51-05 & ~~6 &     & 19.9 & 1601 & ~85 & 0.53 & 0.33 & 10.87 & 12.87 &  100 \\
51-07 & ~~6 &     & 20.2 & 1626 & 112 & 0.67 & 0.32 & 10.99 & 12.80 &  ~65 \\
51-08 & ~23 &  94 & 25.0 & 1552 & 385 & 0.35 & 0.06 & 11.10 & 13.66 &  364 \\
52-01 & ~11 &  29 & 17.1 & 1405 & ~99 & 0.53 & 0.28 & 10.95 & 12.98 &  107 \\
52-02 & ~~8 &     & 13.8 & 1125 & ~75 & 0.64 & 0.45 & 10.91 & 12.77 &  ~73 \\
52-03 & ~~5 &     & 19.1 & 1567 & 105 & 0.45 & 0.23 & 10.58 & 12.86 &  190 \\
52-06 & ~~6 &     & 18.6 & 1531 & ~79 & 0.44 & 0.29 & 10.65 & 12.28 &  ~43 \\
52-07 & ~~6 &     & 23.7 & 1954 & ~48 & 0.44 & 0.49 & 11.07 & 12.25 &  ~15 \\
53-01 & ~13 &  82 & 13.4 & 1006 & 236 & 0.34 & 0.08 & 11.04 & 13.54 &  319 \\
53-03 & ~~5 &     & 15.0 & 1109 & 121 & 0.45 & 0.20 & 10.42 & 13.08 &  460 \\
53-07 & ~15 &  50 & 10.8 & ~833 & ~93 & 0.69 & 0.39 & 10.75 & 12.96 &  161 \\
53-10 & ~~5 &     & 11.4 & ~831 & 103 & 0.45 & 0.23 & 10.51 & 12.76 &  179 \\
54-01 & ~~7 &     & 13.7 & ~802 & 111 & 0.55 & 0.26 & 10.72 & 13.06 &  219 \\
54-03 & ~~5 &     & ~8.4 & ~519 & ~68 & 0.29 & 0.22 & 10.28 & 12.48 &  160 \\
61-11 & ~~6 &     & 23.2 & 1852 & 106 & 0.45 & 0.22 & 10.71 & 12.51 &  ~63 \\
61-16 & ~12 &  40 & 19.3 & 1572 & ~93 & 0.84 & 0.48 & 11.25 & 12.88 &  ~43 \\
65-01 & ~~6 &     & 14.6 & 1132 & ~74 & 0.49 & 0.35 & 10.72 & 12.69 &  ~94 \\
11-24\tablenotemark{a} & ~12 &  92 & 35.2 & 2148 & 311 & 0.69 & 0.12 & 11.74 & 14.00 &  184 \\
31-02\tablenotemark{a} & ~10 &  70 & 37.5 & 2188 & 479 & 0.13 & 0.01 & 11.12 & 13.76 &  434 \\
41-01\tablenotemark{a} & ~11 &  73 & 28.4 & 1807 & 344 & 0.49 & 0.08 & 11.51 & 13.84 &  214 \\

\enddata


 \tablenotetext{a}{Early type group beyond 25 Mpc}


\end{deluxetable}

\clearpage
\tablecaption{Properties of Groups of Dwarf Galaxies. \label{tbl-2}}
\begin{deluxetable}{llccccccccccc}
\rotate
\tabletypesize{\scriptsize}

\tablewidth{0pt}
\tablehead{
\colhead{Group} & \colhead{Principal} & \colhead{No.} & \colhead{Dist.} &
\colhead{$R_I^{3D}$} &
\colhead{$V_r$}  & \colhead{$L_B$} & \colhead{$M_{pm}$} &
\colhead{$M_{\rm v}$}     & \colhead{$M_{pm}/L$}  &
\colhead{$M_{\rm v}/L$}   & \colhead{$M/L_B^{old}$}           &
\colhead{$t_{\rm x} H_0$} \\
 & \colhead{Galaxy} & & \colhead{(Mpc)} & \colhead{(Mpc)} & 
\colhead{(km s$^{-1}$)} & \colhead{($10^8 L_{\odot}$)} & 
\colhead{($10^{11} M_{\odot}$)} & \colhead{($10^{11} M_{\odot}$)} & 
\colhead{($M_{\odot}/L_{\odot}$)} & \colhead{($M_{\odot}/L_{\odot}$)} & 
\colhead{($M_{\odot}/L_{\odot}$)} & 
}
\startdata
14+12 & NGC 3109 & 4 &1.4& 0.34 & 18 & 3.6 & 1.4 & 0.6\tablenotemark{a}&  ~400 &\
~200\tablenotemark{a}& 1220 & 0.83 \\
14~+8 & UGC 8760 & 4 &3.0& 0.29 & 16 & 1.1 & 1.0 &~1.2 & ~860 &1030 & ~250 & 0.76 \\
14+19 & UGC 3974 & 4 &5.0& 0.54 & 28 & 3.3 & 5.1 &~8.5 & 1520 &2560 & 1060 & 0.84 \\
17~+6 & NGC~~~784& 4 &5.0& 0.45 & 36 & 8.5 & 8.3 &10.5 & ~980 &1240 & ~330 & 0.54 \\
\tableline
14+13 & NGC~~~~55& 4 &2.1& 0.30 & 15 &57.6 & 0.9 &~1.1 & ~~16 &\
~~18 & ~~13 & 0.85 \\
      &          & 5\tablenotemark{b} & & 0.47 & 13 &61.5 & 0.8 &~1.1 & ~~14 &\
~~18 &      & 1.5~ \\

\enddata


\tablenotetext{a}{Virial mass estimate biased low: close pair}
\tablenotetext{b}{Including IC 5152 in group}


\end{deluxetable}

\end{document}